\documentclass[12pt]{iopart}
\usepackage{amssymb}
\usepackage[mathscr]{eucal}
\usepackage{epic}
\usepackage{eepic}
\begin{document}
\newtheorem{theorem}{Theorem}
\newcommand{\nc}{\newcommand}
\nc{\ya}{$^1$}\nc{\yb}{$^2$}\nc{\yc}{$^3$}\nc{\yd}{$^4$}\nc{\ye}{$^5$} 
\nc{\beq}{\begin{equation}} \nc{\eeq}{\end{equation}}
\nc{\bea}{\begin{eqnarray}} \nc{\eea}{\end{eqnarray}}
\nc{\ds}{\displaystyle}     \nc{\sy}{\scriptstyle}
\def\EXP{\mbox{{\large\bf e}}}   \nc{\mk}{\makebox}
\nc{\mx}{\mathbf{X}}    \nc{\my}{\mathbf{Y}}    \nc{\mz}{\mathbf{Z}}
\nc{\mX}{\mathfrak{X}}  \nc{\mY}{\mathfrak{Y}}  \nc{\mZ}{\mathfrak{Z}}
\nc{\uop}{\mathbf{u}}  \nc{\wop}{\mathbf{w}}  \nc{\vop}{\mathbf{v}}
\nc{\xop}{\mathbf{x}}  \nc{\zop}{\mathbf{z}}  \nc{\vj}{\mathbf{v}}
\nc{\I}{\mathbf{I}}    \nc{\Vc}{\mathcal{V}}
\nc{\ka}{\kappa}  \nc{\ny}{\nonumber} \nc{\lb}{\langle}  \nc{\rb}{\rangle}
\nc{\lbb}{\langle\!\langle}  \nc{\rbb}{\rangle\!\rangle}
\nc{\al}{\alpha}  \nc{\be}{\beta}     \nc{\ga}{\gamma}   \nc{\de}{\delta}
\nc{\ep}{\epsilon}                    \nc{\sig}{\sigma}
\nc{\lc}{\{}   \nc{\rc}{\}}           \nc{\wf}{\mathfrak{w}}
\nc{\R}{\mathbf{R}}\nc{\Rop}{\mathcal{R}}\nc{\V}{\mathbf{v}\,+\,\mathbf{I}}
\nc{\pnt}{\mathbf{n}}  \nc{\Ropf}{\mathcal{R}^{(f)}} \nc{\Rf}{\mathfrak{R}}
\nc{\one}{\mathbf{e}_1} \nc{\two}{\mathbf{e}_2}  \nc{\thr}{\mathbf{e}_3}
\nc{\F}{\mathbf{F}\,}   \nc{\FF}{\mathfrak{F}\,}
\nc{\lk}{\left(}   \nc{\rk}{\right)}   \nc{\Rb}{\right]}
\nc{\lcb}{\left\{}  \nc{\rcb}{\right\}}  \nc{\Lb}{\left[}
\nc{\dg}{\dagger} \nc{\dgg}{{\dagger\!\dagger}}
\nc{\dgh}{{\dagger\!\dagger\!\dagger}}
\nc{\eeee}[4]{\left[\begin{array}{cc}#1&#2\\#3&#4\end{array}\right]}
\nc{\ee}[2]{\{\overline{#1,#2}\}}
\nc{\vdd}[2]{\vj=\Delta+\int_{#1}^{#2}\omega}
\nc{\thi}[2]
    {\,\Theta\!\left(\!\vj\!+\!\!\!\int_{#1}^{#2}\!\!\!\!\omega\!\right)}
\nc{\thw}[2]{\,\Theta\left(\vj'\!+\!\int_{#1}^{#2}\!\!\!\omega\!\right)}
\nc{\thd}[2]{\,\Theta\left(\Delta\!+\!\!\int_{#1}^{#2}\!\!\omega\!\right)}
\nc{\tha}[2]{\,\Theta\left(\!\Delta\!\!+\!\!\int_{#1}^{#2}\!!\right)}
\nc{\thD}[2]{\,\Theta\left(\Delta\!+\!\int_{#1}^{#2}\;\omega\right)}
\nc{\thn}[2]{\,\Theta\left(\int_{#1}^{#2}\right)}
\nc{\hx}{\hspace{3mm}}     \nc{\hq}{\hspace{6mm}}      \nc{\E}{{\cal{E}}}
\nc{\hs}{\hspace{1cm}}     \nc{\hns}{\hspace*{-9mm}}
\nc{\ok}{\mathbf{e}_k}
\def\sk#1{\scriptstyle{#1}}\def\skk#1{\scriptscriptstyle{#1}}
\def\Xl{{\sf X}}\def\Yl{{\sf Y}}\def\Zl{{\sf Z}}\def\Ul{{\sf U}}
\def\Xt{{\tt X}}\def\Yt{{\tt Y}}\def\Zt{{\tt Z}}\def\Ut{{\tt U}}
\title[Fusion for 3-D integrable Boltzmann weights]{Theta function 
parameterization and fusion for 3-D integrable Boltzmann weights}
\author{G von Gehlen\ya,~S Pakuliak\yb$^,$\yc~and~S Sergeev\yb$^,$\yd$^,$\ye}
\vspace*{4mm}
\address{\ya\ Physikalisches Institut der Universit\"at Bonn,
Nussallee 12, D-53115 Bonn, Germany}
\address{\yb\ Bogoliubov Laboratory of Theoretical Physics,
Joint Institute for Nuclear Research, Dubna 141980, Moscow region, Russia}
\address{\yc\ Institute of Theoretical and Experimental
Physics, Moscow 117259, Russia}
\address{\yd\ Max-Planck-Institut f\"ur Mathematik, Vivatsgasse 7,
D-53111 Bonn, Germany}
\address{\ye\ Dept. of Theoretical 
		Physics, Building 59, Research School of 
		Physical Sciences and Engineering, The Australian National University,
        Canberra ACT 0200, Australia}
\vspace*{2mm}
\eads{\mailto{gehlen@th.physik.uni-bonn.de,}~\mailto{pakuliak@thsun1.jinr.ru,}
 ~\mailto{sergey.sergeev@anu.edu.au}}
\begin{abstract}
We report progress in constructing Boltzmann weights for integrable
3-dimensional lattice spin models. We show that a large class of vertex
solutions to the modified tetrahedron equation can be conveniently
parameterized in terms of $N$-th roots of theta-functions on the Jacobian of
a compact algebraic curve. Fay's identity guarantees the Fermat relations
and the classical equations of motion for the parameters determining the
Boltzmann weights. Our parameterization allows to write a simple formula for
fused Boltzmann weights $\mathfrak{R}$ which describe the partition function of 
an arbitrary open box and which also obey the modified tetrahedron equation.
Imposing periodic boundary conditions we observe that the $\mathfrak{R}$ 
satisfy the normal tetrahedron equation. The scheme described contains the 
Zamolodchikov-Baxter-Bazhanov model and the Chessboard model as special cases.
\end{abstract}\vspace*{-8mm}
\vspace*{-3.6mm}
\vspace*{-4.6mm}
\[\mbox{\small Keywords:~~Tetrahedron equation, 3-D integrable spin models}\]

\section*{Introduction}
The tetrahedron equation  is the  three dimensional
generalization of the Yang-Baxter equation which guarantees
the existence of commuting transfer matrices.
The importance of Yang-Baxter equations for modern mathematics and for
mathematical physics is well known. However, the nature of the tetrahedron
equation is much less understood, this mainly because it is a much more
complicated equation.

Given the physical interest to understand the nature
of the singularities which give rise to 3-D phase transitions, any effort which
gets us closer to analytic results for 3-D statistical systems seems worthwhile.
What has been achieved recently, is to construct large classes of 3-D solvable
models with $\mathbb{Z}_N$-spin variables and to streamline the otherwise
complicated formalism. The goal of an analytic calculation of
partition functions and order parameters is not yet close by. The only
available result from Baxter \cite{bax-part} does not lend itself to
generalizations.

The first solution of the tetrahedron equation was obtained in 1980 by
A. Zamolodchikov \cite{Zamo,Bzam} and then generalized by R. Baxter and 
V. Bazhanov \cite{Bazh-Bax} (ZBB-model) and others \cite{mks}. These
models have $\mathbb{Z}_N$-spin variables and solve the
IRC (Interaction Round a Cube) version of the tetrahedron equation.
Later also the solution of the dual equation, the vertex
tetrahedron equation, was obtained \cite{sms-vertex}, generalizing several
vertex solutions known previously \cite{korepanov,hietarinta}.
Here we shall consider only vertex-type solutions, which are usually denoted by
$\R$, the symmetry will include a $\mathbb{Z}_N$.
In general these $\R$-matrices obey the so called ``simple modified tetrahedron
equation'' which recently has been investigated in \cite{gps}.
The modified tetrahedron equation (MTE) allows us to obtain the ordinary
tetrahedron equation for composite weights or vertices.
In the IRC formulation this has been shown in \cite{mss-mte,bms-mte},
while the most simple vertex case was considered in \cite{ms-mte}.

In this paper we shall introduce a new convenient theta-function
parameterization of general $\R$ operators. This parameterization will allow
us to define fused weights $\Rf$, which are partition functions of open cubes
of size $M^3$, and which obey a MTE. In special cases the $\Rf$ solve an
ordinary tetrahedron equation.

The vertex matrix
$\Rf\in\mathrm{End}\,(\mathbb{C}^{3\,N\,M^2})$ is parameterized
in the terms of $N$-th roots of theta-functions on the Jacobian of a genus
$g=(M-1)^2$ compact algebraic curve $\Gamma_g$.
The divisors of three meromorphic functions on $\Gamma_g$ play the role of
the spectral parameters for $\Rf$. An additional parameter of $\Rf$ is an
arbitrary $\vj\in\mathrm{Jac}(\Gamma_g)$. The tetrahedron equation for $\Rf$
holds due to $M^4$ simple modified tetrahedron equations. In the case when
$M=1$ and therefore $\Gamma_g=S_2$, the solution of the
simple tetrahedron equation of \cite{sms-vertex} is reproduced.

This paper is organized as follows. In section 1 we recall the vertex 
formulation of the 3-D integrable ZBB model and sketch the derivation
of the matrix operator $\R_{ijk}$  from a current
conservation principle and Z-invariance. It satisfies the MTE and can be
parameterized by quadrangle line-sections. In section 2 we introduce the
parameterization of $\R_{ijk}$ in terms of theta functions. The Fermat
relation and the Hirota equations are written as Fay identities.
In section 3 we show that a theta function parameterization allows a
compact formulation of the fusion of many $\R_{ijk}$ to the Boltzmann
weight $\Rf$ of a whole open cube which satisfies a MTE. In section 4
we first consider the special case of vanishing Jacobi transforms,
in which $\Rf$ satisfies a simple TE, then we discuss the rational case and
the relation to Chessboard models. 
Finally, section 5 summarizes our conclusions.

\section{The $\R$-matrix and its parameterization}

In this first section we give a short summary of basic previous results which
also serves for fixing the notation. 

\subsection{The $\R$-matrix of the vertex ZBB model}
We start recalling the vertex formulation of the ZBB-model \cite{sms-vertex}.
We consider a 3-dimensio\-nal lattice with the elementary cell defined by
three non-coplanar vectors $\one,\;\two,\;\thr$ and general vertices
\beq
\pnt\;=\;n_1\,\one+n_2\,\two+n_3\,\thr,\hs \hs n_1,\;n_2,\;n_3\in\:\mathbb{Z}.
\label{cub}  \end{equation}
We label the directed link along $\,\mathbf{e}_j\,$ starting from
$\,\pnt\:$ by $\;(j,\:\pnt).$
On these links there are spin variables $\:\sig_{j,\:\pnt}\:$ which take values
in $\mathbb{Z}_N$. The partition function is defined by
\beq       Z\;=\;\sum_{\{\sig\}}\;\prod_{\pnt}\;
\langle\sig_{1,\pnt},\sig_{2,\pnt+\two},\sig_{3,\pnt}|\:\R\:|
\sig_{1,\pnt+\one},\sig_{2,\pnt},\sig_{3,\pnt+\thr}\rangle\;.
  \label{Zpart}\end{equation}
where $\,\R\,$ is an operator (which in the ZBB model is independent of $\pnt$)
mapping the initial three spin variables to the three final ones, so that
\beq    R^{\sig_1',\sig_2',\sig_3'}_{\sig_1,\sig_2,\sig_3}\;\equiv\;
        \lb\sig_1,\sig_2,\sig_3|\R|\sig_1',\sig_2',\sig_3'\rb
\label{RMa}\end{equation}
is a $\:N^3\times N^3\:$ matrix independent of $\pnt$.


For the vertex ZBB-model, (\ref{RMa}) can be expressed as a kind of cross
ratio of four cyclic functions $W_p(n)$.
Introduce a two component vector $p=(x,y)$ which is restricted to the Fermat
curve
\beq x^N\,+\,y^N\,=\,1\,.\label{ferma}\end{equation}
Then define the function $W_p(n)$ by
\beq  W_p(0)\;=\;1,\hs W_p(n)\;=\;\prod_{\nu=1}^n\;\frac{y}{1\,-\,q^\nu\,x}
  \hspace{5mm} \mbox{for}\;\;n>0.            \label{W-def}     \end{equation}
where \\[-9mm]
\beq q\;=\;e^{2\pi\,i/N} \label{qN} \end{equation}
is the primitive $N$$-th$ root of unity.
Because of the Fermat curve restriction, $W_p(n)$ is cyclic in $\,n\,$:
$$W_p(n+N)\;=\;W_p(n).$$
Now $\R\,=\,\R(p_1,p_2,p_3,p_4)$ is defined by the following matrix function
depending on four Fermat points $\:p_1,p_2,p_3,p_4$
\begin{equation}\label{R-matrix}
 R^{\sig_1',\sig_2',\sig_3'}_{\sig_1,\sig_2,\sig_3}
\;\stackrel{\textrm{\it\footnotesize def}}{=}\;
\delta_{\sig_2+\sig_3,\sig'_2+\sig'_3}\;q^{(\sig'_1-\sig_1)\,\sig'_3}\;
\frac{W_{p_1}(\sig_2-\sig_1)\,W_{p_2}(\sig'_2-\sig'_1)}
{W_{p_3}(\sig'_2-\sig_1)\,W_{p_4}(\sig_2-\sig'_1)}\;,
\end{equation}
where $x$-coordinates of four Fermat curve points in (\ref{R-matrix})
are identically related by
\begin{equation}\label{xxxx}
\ds x_1^{}\;x_2^{}\;=\; q\;\;x_3^{}\;x_4^{}\;.
\end{equation}
So the matrix elements
$\:R^{\sig_1',\sig_2',\sig_3'}_{\sig_1,\sig_2,\sig_3}\:$
depend on three complex numbers. These correspond to 
Zamolodchikov's spherical angles in the IRC-formulation of the 
ZBB-model \cite{sms-vertex}.
The structure of the indices of the matrix (\ref{R-matrix})
allows one to consider $\R$ as the operator acting in the tensor
product of three vector spaces
\begin{equation}\label{Vspace}
\mathcal{V}\;=\;\mathbb{C}^N\;,\hs\hs
\R\;\in\;\mathrm{End}\left(\mathcal{V}\otimes\mathcal{V}\otimes\mathcal{V}
\right)\,. \end{equation}
It is conventional to enumerate naturally the components of the tensor product
of several vector spaces, so that (\ref{R-matrix}) are
the matrix elements of $\R\;=\;\R_{123}$. Of course,
$\R_{123}$ acts trivially on the all other vector spaces
if one considers $\mathcal{V}^{\otimes \Delta}$ for some arbitrary $\Delta$.

\eref{R-matrix} is known as the $\R$-matrix of the
Zamolodchikov-Bazhanov-Baxter model, see \cite{sms-vertex}.
The proof that (\ref{R-matrix}) satisfies the Tetrahedron Equation
\bea \label{bbzmte-el}\lefteqn{\sum_{j_1...j_6}\;
R_{i_1i_2i_3}^{j_1j_2j_3}(p^{(1)})\,R_{j_1i_4i_5}^{k_1j_4j_5}(p^{(2)})
\,
R_{j_2j_4i_6}^{k_2k_4j_6}(p^{(3)})\,R_{j_3j_5j_6}^{k_3k_5k_6}(p^{(4)})}
\ny\\[-2mm] \hq\;&=&\sum_{j_1...j_6}\;
R_{i_3i_5i_6}^{j_3j_5j_6}(p^{(4)})\,R_{i_2i_4j_6}^{j_2j_4k_6}(p^{(3)})
\,
R_{i_1j_4j_5}^{j_1k_4k_5}(p^{(2)})\,R_{j_1j_2j_3}^{k_1k_2k_3}(p^{(1)})
\, \label{ZBBtetr}\eea
is rather tedious \cite{sms-vertex}. In (\ref{ZBBtetr}) the arguments
$\,p^{(j)}\;(j=1,\ldots,4)\,$ stand for four Fermat curve points
$\:(p_1^{(j)},p_2^{(j)},p_3^{(j)},p_4^{(j)})\:$ each. These 16 points depend
on five independent parameters expressible in terms of spherical angles,
see \cite{sms-vertex}. Note that here on the left and right hand side the
same $p^{(j)}$ appear. This will not be the case in the generalizations which
will be discussed soon.

Baxter and Forrester \cite{bx-forr-crit} have studied whether this model
describes phase transitions. They used variational and numerical methods and
found strong evidence that for the parameter values for which (\ref{ZBBtetr})
is satisfied, the ZBB-model is just at criticality \cite{bx-forr-crit}.
So, in order to get a chance to describe phase transitions while staying
integrable (recall that also for the 2D Potts model this is a problem),
one should enlarge the framework and define more general Boltzmann weights and
introduce less restrictive Tetrahedron equations. Less restrictive and still
powerful generalized equations can be used, as has been shown by Mangazeev
and Stroganov \cite{mss-mte}: they introduced Modified Tetrahedron Equations
which guarantee commuting layer-to-layer transfer matrices.
Further work along this line has been done in \cite{bms-mte,ms-mte}.

\subsection{$\;\R$-matrix satisfying the Modified Tetrahedron Equation}

Since in the above-mentioned work \cite{sms-vertex} the proof that particular 
Boltzmann weights satisfy a particular MTE has been rather tedious, here we 
shall follow the approach introduced in \cite{s-qem} 
in which there is no need for an explicit check of the MTE. 
The Boltzmann weights are constructed from "physical"
principles which guarantee the validity of the MTE and nevertheless leave
much freedom to obtain a broad class of integrable 3D-models. We give a short
summary of the argument.

One starts with an {\it oriented} 3-D basic lattice. The dynamic variables
living on the links $i$ of this lattice are taken to be elements
$\,u_i,w_i\in\wf_i\,$ an ultralocal Weyl algebra $\wf\:=\:\bigotimes \wf_i$
at the primitive $N$-$th$ root of unity:
$\;u_i\,w_j\;=\;q^{\delta_{ij}}\,w_j\,u_i\,,$ ($q$ as in
eq.(\ref{qN})), which generalize the $\mathbb{Z}_N$ spin variables
of (\ref{Zpart}) and (\ref{RMa}). The Weyl elements are represented by
standard $N\times N$  raising respectively diagonal matrices.
The $N$-$th$ powers of the Weyl variables are centers of the algebra and so
are scalar variables.

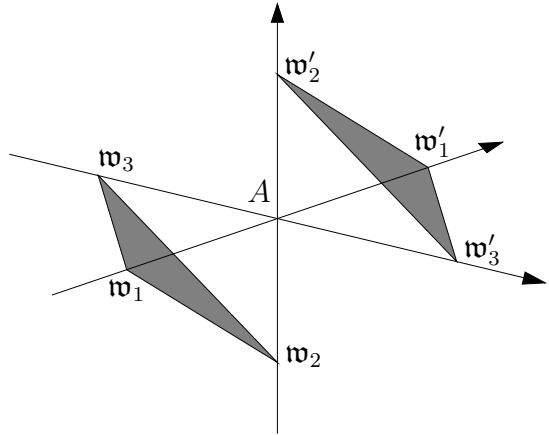
\begin{figure}[ht]
\begin{center}
\renewcommand{\dashlinestretch}{100}
\unitlength=0.07pt
\begin{picture}(3300,2150)(2000,1600)
\shade\path(2648, 2222)(2493, 2732)(3462, 1721)(2648, 2222)
\shade\path(4277, 2776)(3462, 3277)(4431, 2266)(4277, 2776)
\path(4678, 2912)(3462, 2499)
\path(3462, 1338)(3462, 2499)(2016, 2847)
\path(3462, 2499)(3462, 3660)
\path(2246,2086)(3462,2499)(4908,2151)
\put(2548,2080){$\wf_1$} \put(2493,2782){$\wf_3$}
\put(3505,1721){$\wf_2$}
\put(4200,2870){$\wf_1'$}\put(3500,3277){$\wf_2'$}
\put(4471,2290){$\wf_3'$}
\put(3300,2585){$A$}
\blacken\path(3430,3560)(3462,3660)(3494,3560)(3430,3560)  
\blacken\path(4785,2145)(4908,2151)(4800,2210)(4785,2145)  
\blacken\path(4595,2855)(4678,2912)(4550,2905)(4595,2855)  
\end{picture}
\end{center}
\caption{\footnotesize{The six links of the basic lattice intersecting in
the vertex A, intersected by auxiliary planes (shaded) in two different
positions: first passing through $\wf_1,\,\wf_2,\,\wf_3$ and second
through $\wf_1',\,\wf_2',\,\wf_3'$. The second position is obtained from the
first by moving the auxiliary plane parallel through the vertex $A$. The Weyl
variables, elements of $\wf_i,\,\wf_i'$ live on the links of the basic lattice.
$\Rop$ maps the left auxiliary triangle onto the upper right one.}}
\label{tri}\end{figure}

The main object constructed is an invertible canonical mapping
$\Rop_{ijk}$ in the space of a triple Weyl algebra. $\Rop_{ijk}$
operates at the vertices of the 3D lattice, mapping the three Weyl elements
on the "incoming" links onto those on the "outgoing" links,
see \Fref{tri}.
%
%
\begin{figure}[ht]
\setlength{\unitlength}{0.5mm}
\begin{center}
\begin{picture}(300,120)
{\large \put( 50,  0){\vector( 0,1){110}} \put(230,0){\vector( 0,1){110}}
\put( 10, 20){\vector( 2,1){110}}
\put(160, 40){\vector( 2,1){110}}
\put(120, 40){\vector(-2,1){110}}
\put(270, 20){\vector(-2,1){110}}
\put(  5, 84){\mk(0,0)[lb]{$\Zl$}}   \put(160, 80){\mk(0,0)[lb]{$\Zl$}}
\put( 55,110){\mk(0,0)[lb]{$\Yl$}}   \put(222,110){\mk(0,0)[lb]{$\Yl$}}
\put(110, 80){\mk(0,0)[lb]{$\Xl$}}   \put(270, 85){\mk(0,0)[lb]{$\Xl$}}
\put( 52, 75){\mk(0,0)[lb]{$\wf_1$}}
\put(218, 75.5){\mk(0,0)[lb]{$\wf_3'$}}
\put( 52, 33){\mk(0,0)[lb]{$\wf_3$}}
\put(218, 31.0){\mk(0,0)[lb]{$\wf_1'$}}
\put( 90.5, 53.5){\mk(0,0)[lb]{$\wf_2$}}
\put(178.5, 54.2){\mk(0,0)[lb]{$\wf_2'$}}}
\put( 52, 14){\mk(0,0)[lb]{$c_1$}}
\put(222, 14){\mk(0,0)[lb]{$c_1$}}
\put( 52, 97){\mk(0,0)[lb]{$c_3$}}
\put(222, 97){\mk(0,0)[lb]{$c_3$}}
\put( 30, 34){\mk(0,0)[lb]{$b_1$}}
\put(250, 31){\mk(0,0)[lb]{$d_1$}}
\put( 30, 87){\mk(0,0)[lb]{$d_3$}}
\put(250, 89){\mk(0,0)[lb]{$b_3$}}
\put( 42, 55){\mk(0,0)[lb]{$c_2$}}
\put(233, 55){\mk(0,0)[lb]{$c_2'$}}
\put(204, 41){\mk(0,0)[lb]{$d_2'$}}
\put( 71, 43){\mk(0,0)[lb]{$b_2$}}
\put( 70, 66){\mk(0,0)[lb]{$d_2$}}
\put(204, 66){\mk(0,0)[lb]{$b_2'$}}
\put(100, 39){\mk(0,0)[lb]{$d_1$}}
\put(169, 36.5){\mk(0,0)[lb]{$b_1$}}
\put(172, 69){\mk(0,0)[lb]{$d_3$}}
\put( 98, 69){\mk(0,0)[lb]{$b_3$}}
\put(127,109){\mk(0,0)[lb]{\large$\Rop_{1\,2\,3}$}}
\put(127,100){\mk(0,0)[lb]{\LARGE$\longrightarrow$}}
\end{picture}
\caption{\footnotesize{The canonical invertible mapping $\Rop_{123}$ shown in
the auxiliary planes passing through the incoming (left) and outgoing (right)
dynamic variables which are elements of $\wf_i$ resp. $\wf_i'$. The directed
lines $X,\:Y,\:Z$ are the intersections of the three planes forming the vertex
$A$ of \Fref{tri}. Their sections are labeled by the line-section
parameters $b_1,\:,\ldots,\:d_3$. Note that the choice of the orientation of 
the lines is not unique. The orientation chosen here corresponds to the 
numbering (\ref{inisig}) and (\ref{outsig}) of the fused vertex considered 
in the \Sref{section3}.}}
\label{fumap}\end{center}\end{figure}
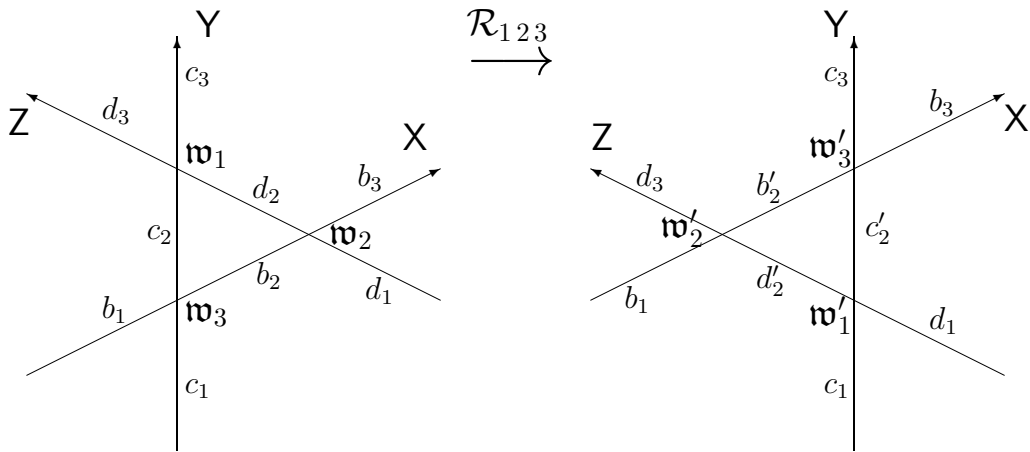
The construction of
$\Rop_{ijk}$ is based on two postulates, a Kirchhoff-like current
conservation and a Baxter Z-invariance, and gives a unique explicit result: a
canonical and invertible rational mapping operator. Since $q$ is a root of
unity, $\Rop_{ijk}$ decomposes into a matrix conjugation $\R_{ijk}$, and a
purely functional mapping $\Ropf_{ijk}$ which acts on the scalar parameters
(the Weyl centers).
So, for any rational function $\Phi$ on $\wf$:
\beq \Rop_{123}\circ\Phi\;=\;\R_{123}(\Ropf_{123}\circ\Phi)\R_{123}^{-1}\,.
      \label{decomp}\end{equation}
It turns out that the matrix $\R_{ijk}$ has the form (\ref{R-matrix})
where the four Fermat curve parameters, again constrained by (\ref{xxxx}),
are rational functions of the scalar Weyl center parameters.

Next consider an auxiliary plane which cuts the three incoming links near a
vertex, and a second auxiliary plane cutting through the outgoing links, see
\Fref{tri}. We take the six Weyl dynamic variables to sit on the six
intersection points of the auxiliary planes.
$\Rop_{123}$ can be regarded as the mapping of the ingoing auxiliary plane
to the parallel shifted outgoing auxiliary plane.

Now consider the vertices of the basic lattice to be formed as the
intersection points of three sets of non-parallel planes.
The three planes which form the vertex $A$ of \Fref{tri} intersect the
auxiliary planes in the lines $\:\Xl,\;\Yl,\;\Zl\:$ shown in \Fref{fumap}.
In \Fref{tri} these intersection lines are the sides of the shaded triangles.
Seen from the moving auxiliary plane, $\Rop_{123}$ shifts the line $X$
through the vertex with index $1$ or $Y$ through the vertex $2$ etc. We attach
variables $b_1,\:b_2,\ldots,\:d_3$ to each section of the lines
$\,\Xl,\:\Yl,\:\Zl\,$
as shown in \Fref{fumap}.

It is convenient to parameterize the two scalar variables associated with
the incoming dynamic Weyl variable $\wf_1$ (corresponding to $u_1^N$
 and $w_1^N$ in usual notation) by the ratios $c_2^N/c_3^N$ and
$d_3^N/d_2^N$. Analogously, e.g. those for $\wf_2'$ are defined as
$b_1^N/{b_2'}^N$ and ${d_2'}^N/d_1^N$ etc. Details of the rule to parameterize
the scalar variables in terms of "line-section" variables $b_1,\ldots,d_3$ etc.
are explained in \cite{gps}. However, these will not be essential here,
since one of the aims of this paper is to introduce and use another
parameterization. Just observe in \Fref{fumap} that $\Ropf_{123}$
changes only three of the line-section parameters: $b_2,\:c_2,\:d_2$.
From the explicit form of the canonical operator $\Rop_{123}$ (see \cite{gps})
one finds that the functional mapping $\Ropf$ is rational in the $N$-$th$
powers of the line sections:
\bea
{b_2}'^N&=&\frac{b_1^Nc_3^Nd_2^N+b_2^Nc_3^Nd_3^N
+\ka_1^Nb_3^Nc_2^Nd_3^N}{c_2^Nd_2^N };\ny\\
{c_2}'^N&=&\frac{\ka_1^Nb_3^Nc_1^Nd_2^N+\ka_3^Nb_2^Nc_3^Nd_1^N
+\ka_1^N\ka_3^Nb_3^Nc_2^Nd_1^N}{\ka_2^Nb_2^Nd_2^N};\ny\\
{d_2}'^N&=&\frac{b_2^Nc_1^Nd_3^N+b_1^Nc_1^Nd_2^N
+\ka_3^Nb_1^Nc_2^Nd_1^N}{b_2^Nc_2^N}.       \label{fli}\eea
Here $\:\ka_1,\:\ka_2,\:\ka_3\:$ are fixed parameters ("coupling constants")
of the mapping $\Rop_{123}$. One can show \cite{gps} that
in the line-section parameterization the three independent Fermat curve
parameters which determine $\R_{123}$ according to (\ref{R-matrix}),(\ref{xxxx})
 are
\beq x_1=\frac{b_2c_3}{\ka_1b_3c_2};\hs x_2=\frac{\ka_2b_1c_2'}{b_2'c_1};
\hs x_3=\frac{b_1c_3}{\sqrt{q}\:b_2'c_2}.\label{fuei}\end{equation}
If we define a partition function in analogy to (\ref{Zpart}),
the matrix elements of $\R_{ijk}$ take the role of generalized (because 
generically they will be complex) Boltzmann weights of integrable 3D lattice 
models of statistical mechanics. Despite their non-positivity we shall just call
these matrix elements ``Boltzmann weights".

Via the physical assumptions made in constructing $\Rop_{ijk}\,,$ 
the validity of
the TE is already built in. Simply considering two different sequences of 
Z-invariance shifts in a geometric figure formed by {\it four} intersecting
straight lines ("quadrangle"), one concludes that (see e.g. \cite{gps})
\beq\label{op-te}
\ds \Rop_{123}\cdot\Rop_{145}\cdot\Rop_{246}\cdot\Rop_{356}\;\sim\;
\Rop_{356}\cdot\Rop_{246}\cdot\Rop_{145}\cdot\Rop_{123}\,,\label{tetra}
\end{equation}
i.e. that $\Rop_{ijk}$ satisfies the Tetrahedron equation.
Inserting (\ref{decomp}) into (\ref{op-te}) and choosing various phases
of $N$-$th$ roots (the Fermat points \eref{fuei} involve $b_1,\ldots$, 
whereas the \eref{fli} relate $b_1^N,\ldots$), leads to the MTE for the matrix 
operator $\,\R_{ijk}$:
\beq\label{mte-general}\fl
\ds\begin{array}{l}\ds\R_{123} \cdot\lk\Ropf_{123}\circ\R_{145}\rk
\!\cdot\!\lk\Ropf_{123}\Ropf_{145}\circ\R_{246}\rk
\!\cdot\!\lk\Ropf_{123}\Ropf_{145}\Ropf_{246}\circ\R_{356}\rk
\\[4mm]\hs
\sim\; \R_{356}\cdot\lk\Ropf_{356}\circ\R_{246}\rk
\!\cdot\!\lk\Ropf_{356}\Ropf_{246}\circ\R_{145}\rk
\!\cdot\!\lk\Ropf_{356}\Ropf_{246}\Ropf_{145}\circ\R_{123}\rk\,.
\end{array}\end{equation}
Via the Fermat points each $\R_{ijk}$ depends on several scalar variables,
see e.g. \eref{fuei}. In \eref{mte-general} the scalar variables which appear 
in the matrices $\R_{ijk}$ are to be transformed by the functional 
transformations $\Ropf_{ijk}$ as indicated.
Let us write shorthand
\bea \fl\R^{(1)}&=&
\R_{123};\hs\hs\hs\hx\;\,\R^{(2)}\;=\;\Ropf_{123}\circ\R_{145};\hs\hx\:
\R^{(3)}\;=\;\Ropf_{123}\Ropf_{145}\circ\R_{246};\ny\\[1mm] \fl
\R^{(4)}&=&\Ropf_{123}\Ropf_{145}\Ropf_{246}\circ\R_{356};\hq
\R^{(5)}\;=\;\Ropf_{356}\Ropf_{246}\Ropf_{145}\circ\R_{123};\ny\\[1mm] \fl
\R^{(6)}&=&\Ropf_{356}\Ropf_{246}\circ\R_{145};\hs\hx\;
\R^{(7)}\;=\;\Ropf_{356}\circ\R_{246};\hs\hx\; \R^{(8)}\;=\;\R_{356}.
   \label{teab}\eea
Then (\ref{tea}) becomes
\beq \R^{(1)}\,\R^{(2)}\,\R^{(3)}\,\R^{(4)}\;=\;\rho\;
\R^{(8)}\,\R^{(7)}\,\R^{(6)}\,\R^{(5)},  \label{rff} \end{equation}
where each $\,\R^{(j)}\:$ acts non-trivially in only three of the six spaces
$\:\mathcal{V}\:=\:\mathbb{C}^N.$
$\rho$ is a scalar density factor which comes in when passing from mappings to
matrix equations.

The parameters which determine the $\R^{(j)}$ are the corresponding Fermat
curve coordinates. Taking into account the functional transformations in
(\ref{teab}) in terms of the line-section parameters one finds (for full details
see \cite{gps}):
$\;\R^{(j)}\:=\:\R(x_1^{(j)},x_2^{(j)},x_3^{(j)})\;$ with
\beq \left(\begin{array}{ccc}
  x_1^{(1)}&x_2^{(1)}&x_3^{(1)}\\[3mm] x_1^{(2)}&x_2^{(2)}&x_3^{(2)}\\[3mm]
   \vdots&\vdots&\vdots\\[3mm]
  x_1^{(7)}&x_2^{(7)}&x_3^{(7)}\\[3mm] x_1^{(8)}&x_2^{(8)}&x_3^{(8)}\end{array}
      \right)\;=\;q^{-1/2}\:\left(\begin{array}{ccc}
\ds\frac{b_2c_3}{\ka_1b_3c_2}&\ds\frac{\ka_2b_1c_2'}{b_2'c_1}&
\ds\frac{b_1c_3}{q^{1/2}\:b_2'c_2}\\
\ds\frac{a_2c_2'}{\ka_1a_3c_1}&\ds\frac{\ka_4a_1c_1''}{a_2''c_0}&
\ds\frac{a_1c_2'}{q^{1/2}\:a_2''c_1}\\ \vdots&\vdots&\vdots\\
\ds\frac{a_2b_3}{\ka_2a_3b_2}&
\ds\frac{\ka_4a_1^\dg b_2^\dgg}{a_2^\dgg b_1^\dg}&
\ds\frac{a_1^\dg b_3}{q^{1/2}\:a_2^\dgg b_2}\\
\ds\frac{a_1 b_2}{\ka_3a_2b_1} &
\ds\frac{\ka_5a_0b_1^\dg}{b_0a_1^\dg} &
\ds\frac{a_0b_2}{q^{1/2}\:a_1^\dg b_1}\end{array}\right).\label{xa}
\end{equation}
The once or multiply transformed parameters like $c_1'',\;b_1^\dg$
follow from the iteration of equations like (\ref{fli}). Altogether, since
there are eight matrices $\R^{(j)}$ appearing in the MTEs, and as seen in
\Fref{fumap}, each transformation changes three line-section parameters,
we have 24 equations for 32 different line-section
parameters (these parameters can seen in \Tref{abcdt} below). This is a
set of classical integrable equations which conveniently
are written in Hirota form:
\bea
{b_2}'^Nc_2^Nd_2^N&=&b_1^Nc_3^Nd_2^N+b_2^Nc_3^Nd_3^N
+\ka_1^Nb_3^Nc_2^Nd_3^N;\ny\\
\ka_2^Nb_2^Nc_2'^Nd_2^N&=&\ka_1^Nb_3^Nc_1^Nd_2^N+\ka_3^Nb_2^Nc_3^Nd_1^N
+\ka_1^N\ka_3^Nb_3^Nc_2^Nd_1^N;\ny\\
b_2^Nc_2^N{d_2}'^N&=&b_2^Nc_1^Nd_3^N+b_1^Nc_1^Nd_2^N
+\ka_3^Nb_1^Nc_2^Nd_1^N;\ny\\[3mm]
a_2''^Nc_1^Nd_1^N&=&a_1^Nc_2'^Nd_1^N+a_2^Nc_2'^N{d_2}'^N
+\ka_1^Na_3^Nc_1^N{d_2}'^N;\ny\\
\ka_4^Na_2^Nc_1''^Nd_1^N&=&\ka_1^Na_3^Nc_0^Nd_1^N+\ka_5^Na_2^Nc_2'^Nd_0^N
+\ka_1^N\ka_5^Na_3^Nc_1^Nd_0^N;\ny\\
a_2^Nc_1^Nd_1''^N&=&a_2^Nc_0^N{d_2}'^N+a_1^Nc_0^Nd_1^N
+\ka_5^Na_1^Nc_1^Nd_0^N;\ny\\[3mm]
 a_1'''^Nb_1^N{d_2}'^N&=&a_0^N{b_2}'^N{d_2}'^N+d_3^Na_1^N{b_2}'^N
+\ka_2^Nd_3^Nb_1^Na_2''^N;\ny\\
 \ka_4^Na_1^Nb_1'''^N{d_2}'^N&=&\ka_2^Nb_0^Na_2''^N{d_2}'^N
+\ka_6^Na_1^N{b_2}'^Nd_1''^N
 +\ka_2^N\ka_6^Na_2''^Nb_1^Nd_1''^N;\ny\\
 a_1^Nb_1^N{d_2}'''^N&=&b_0^Nd_3^Na_1^N+a_0^Nb_0^N{d_2}'^N
+\ka_6^Na_0^Nb_1^Nd_1''^N;\ny\\[3mm]
 {a_2^\dgg}^N{b_2}'^Nc_2'^N&=&b_3^Na_1'''^Nc_2'^N+b_3^Nc_3^Na_2''^N
+\ka_3^Na_3^Nc_3^N{b_2}'^N;\ny\\
 \ka_5^Na_2''^N{b_2^\dgg}^Nc_2'^N&=&\ka_3^Na_3^Nb_1'''^Nc_2'^N
+\ka_6^Nb_3^Na_2''^Nc_1''^N
  +\ka_3^N\ka_6^Na_3^N{b_2}'^Nc_1''^N;\ny\\
 a_2''^N{b_2}'^N{c_2^\dgh}^N&=&c_3^Na_2''^Nb_1'''^N+a_1'''^Nb_1'''^Nc_2'^N
+\ka_6^Na_1'''^N{b_2}'^Nc_1''^N;
\ny \\[1mm]
\vdots && \vdots\ny \\[2mm]
{b_1'''}^N{c_1^\dg}^N {d_1^{\dgg}}^N&=&b_0^N{c_2^{\dgh}}^N{d_1^{\dgg}}^N
+{b_1^\dg}^N {c_2^{\dgh}}^N{d_2^{\dgh}}^N
             +\ka_1^N{b_2^{\dgg}}^N{c_1^\dg}^N {d_2^{\dgh}}^N;\ny\\
\ka_2^N{b_1^\dg}^N {c_1''}^N{d_1^{\dgg}}^N&=&
\ka_1^Nc_0^N{b_2^{\dgg}}^N{d_1^{\dgg}}^N
          +\ka_3^Nd_0^N{b_1^\dg}^N {c_2^{\dgh}}^N
          +\ka_1^N\ka_3^Nd_0^N{b_2^{\dgg}}^N{c_1^\dg}^N;\ny\\
{b_1^\dg}^N {c_1^\dg}^N {d\,''_{\!1}}^N&=&
c_0^N{b_1^\dg}^N {d_2^{\dgh}}^N+b_0^Nc_0^N{d_1^{\dgg}}^N
+\ka_3^Nb_0^Nd_0^N{c_1^\dg}^N.
\label{hiro}\eea
The first three of these equations are just (\ref{fli}), defining $\Ropf_{123}$,
i.e. ${b_2'}^N,\;{c_2'}^N,\;{d\,'_{\!2}}^N$ in terms of the unprimed 
$b_1,\:\ldots,\:d_3\,.\;$ The first six equations
together (e.g. express in the fourth eq. on the right hand side $c_2'^N$ and
$d_2'^N$ from the first and third equations)
define $\Ropf_{123}\circ\Ropf_{145}$,
etc. The complete expressions for (\ref{xa}) and (\ref{hiro}) can be found
in \cite{gps}.

Straightforward combination of the first twelve 
equations of (\ref{hiro}) on one hand, and of the last twelve of equations 
of (\ref{hiro}) on the other hand (best done e.g. by Maple), 
shows that the functional mappings given 
in (\ref{hiro}) automatically satisfy the functional TE:
\beq \Ropf_{123}\cdot\Ropf_{145}\cdot\Ropf_{246}\cdot\Ropf_{356}
     \;=\;\Ropf_{356}\cdot\Ropf_{246}\cdot\Ropf_{145}\cdot\Ropf_{123}
\label{tea}\end{equation}
where for the superposition of two operators acting on a function $\Phi$ we
use the notation
\mbox{$(({\cal A}\cdot{\cal B})\cdot\Phi)\stackrel{def}{=}\;
          ({\cal A}\cdot ({\cal B}\cdot\Phi))$.}
Of course, the validity of (\ref{tea}) is a consequence of the physical rules
used when constructing $\Rop_{ijk}$. In the line-section parameterization
the relation between the first, second etc. lines in both (\ref{xa}) and in
(\ref{hiro}) is not transparent. Introducing a new parameterization in the
next subsection will make these relations simple and explicit.

\section{Parameterization using concepts of algebraic geometry}
\subsection{Theta functions}

It is well-known \cite{krich,Shi,Mul,KWZ,ser-exact} that Hirota-type equations
can be identically satisfied by a parameterization in terms of theta functions
on an algebraic curve. We shall now introduce such a parameterization in order
to write \eref{xa} and \eref{hiro} in a more systematic way. This will be useful
also later to formulate fusion in a transparent manner.
For the notations of algebraic geometry see e.g. \cite{mumford}.

Let $\Gamma_g$ be an abstract generic algebraic curve of the genus $g$
with \mbox{\boldmath{$\omega$}} being
the canonical $g$-dimensional vector of the homomorphic differentials.
For any two points $X,Y\in\Gamma_g$ let
$\;\;\I_Y^X\;:\;\Gamma_g^2\mapsto\mbox{Jac}(\Gamma_g)\;$ be
\beq \I_Y^X\;\stackrel{\textrm{\it\footnotesize def}}{=}\;
   \int_Y^X\,\mbox{\boldmath{$\omega$}}\;.\end{equation}
Let further $\:E(X,Y)\:=-\,E(Y,X)\:$ be the prime form on $\Gamma^2_g\,$, and
$\,\Theta(\vj)\,$ be the theta-function on $\,\mathrm{Jac}(\Gamma_g)$.

It is well known, the theta-functions on the Jacobian of an algebraic
curve obey the Fay identity
\bea \fl \Theta(\vj)\;\Theta(\V_B^A\,+\,\I_D^C)&\;\;=\;\;&
 \Theta(\V_D^A)\;\Theta(\V_B^C)\:\frac{E(A,B)\:E(D,C)}{E(A,C)\:E(D,B)}
 \ny\\ \fl &\;\;+\;\;&
 \Theta(\V_B^A)\;\Theta(\V_D^C)\:\frac{E(A,D)\:E(C,B)}{E(A,C)\:E(D,B)}\;,
 \label{HI}\end{eqnarray}
which involves four points $A,B,C,D\in\Gamma_g$ and a $\vj\in
\mbox{Jac}(\Gamma_g).$
We shall show that in the parameterization to be introduced below, the Fermat
relations become just Fay-identities. The Fay identity involves only
cross ratios of prime forms, and these ratios
 have a simple expression in terms of
non-singular odd characteristic theta functions:
\beq \eeee{X}{X'}{Y}{Y'}\:\:\stackrel{\textrm{\it\footnotesize def}}{=}\:\:
  \frac{E(X,Y)\,E(X',Y')}{E(X,Y')\,E(X',Y)}\:=\:
  \frac{\Theta_{\ep_{odd}}(\:\I_X^Y\:)\:\Theta_{\ep_{odd}}(\:\I_{X'}^{Y'}\:)}
  {\Theta_{\ep_{odd}}(\:\I_X^{Y'}\:)\:\Theta_{\ep_{odd}}(\:\I_{X'}^Y\:)}\
  .       \label{crossr}\end{equation}
For solving the 24 trilinear equations \eref{hiro} we
shall need an identity with more arguments $Q,X,Y,Y',Z,Z'\in \Gamma_g$ 
obtained by combining two Fay identities:
\bea \fl
\lefteqn{\Theta(\vj+\I^Q_X)\,\Theta(\vj+\I^Q_Y+\I^{Z'}_Z)\,
   \Theta(\vj+\I^Q_Z+\I^{Y'}_Y)} \ny \\ \fl \hq &-&\;
\Theta(\vj+\I^Q_X+\I^{Z'}_Z)\,\Theta(\vj+\I^Q_Y)\,
       \Theta(\vj+\I^Q_Z+\I^{Y'}_Y)\,
\eeee{X}{Y}{Z}{Z'}\ny\\ \fl \hq &-&\;
\Theta(\vj+\I^Q_X+\I^{Y'}_Y)\,\Theta(\vj+\I^Q_Y+\I^{Z'}_Z)\,\Theta(\vj+\I^Q_Z)\,
\eeee{X}{Z}{Y}{Y'}\ny\\  \fl \hq &+&\;
\Theta(\vj+\I^Q_X+\I^{Y'}_Y+\I^{Z'}_Z)\,\Theta(\vj+\I^Q_Y)\,\Theta(\vj+\I^Q_Z)\,
\eeee{X}{Y}{Z}{Z'}\;\eeee{X}{Z'}{Y}{Y'}\;=\;0\,.
\label{ttt} \eea
Furthermore, since we will need the $N$-th roots of theta-functions and prime 
forms, we also define $e(X,Y)$ and $\theta(\vj)$ by
\beq  \label{etheta}
e(X,Y)^N\;=\;\Theta_{\ep_{odd}}(\I^X_Y)\;\sim\;E(X,Y)\:;\hs
\theta(\vj)^N\;=\;\Theta(\vj)\:.
\end{equation}
Since in the following we shall have to write many equations involving
theta-functions, it is convenient to introduce special abbreviations.
For  $\;Q,A,B_1,B_1',\ldots\in \Gamma_g\;$ we define:
\begin{eqnarray} \fl
(A,\:B_1+B_2+\ldots+B_n)&\equiv&
    \Theta(\vj+\I^Q_A+\sum_{j=1}^n\I^{{B_j}'}_{B_j})\,;\hs
        \lb A,B\rb \;\equiv\; E(A,B)\,;\ny\\[-1mm] \fl
 \left[A, \:B_1+B_2+\ldots + B_n\:\right]&\equiv&
\theta(\vj+\I^Q_A+\sum_{j=1}^n\I^{{B_j}'}_{B_j})\,;
\ny\\[-1mm]  \fl  \lc A,B\rc\;\equiv\;-\,q^{-1/2}\,e\,(A,B')\,/\,e(A,B)\,;
     \hspace{-22mm}&&\hs\hs\hs
\ee{A}{B}\;\equiv\;-e(A',B)\,/\,e(A,B)\,.   \label{defth}\end{eqnarray}
Note that the brackets $\,(\;,\;)\,$ and $\,\left[\;,\;\right]\,$ 
introduced here do not show explicitly the
dependence on the variables $\,\vj,\,Q,\,B_1',\ldots B_n'\,$ since these
always come in the same form. \\[2mm]
We also introduce, using these notations:
\bea \fl \lefteqn{\F(\vj;\,X,Y',Y,Z',Z)\:\: 
\stackrel{\textrm{\it\footnotesize def}}{=}
		 \;(X)(Y,Z)(Z,Y)\;\lb Y,Z\rb\,\lb X,Z'\rb\,\lb X,Y'\rb} \ny\\[2mm] \fl
\hq&-& (X,Z)(Y)(Z,Y)\;\lb X,Z\rb\,\lb Y,Z'\rb\,\lb X,Y'\rb 
 \;-\;(X,Y)(Y,Z)(Z)\;\lb X,Y\rb \,\lb Y',Z\rb\,\lb X,Z'\rb \ny\\[2mm] \fl
\hq &+&(X,Y+Z)(Y)(Z)\;\lb X,Z\rb \,\lb X,Y\rb\,\lb Y',Z' \rb\,,  \label{DFay}
\end{eqnarray}
so that the Double-Fay-identity (\ref{ttt}) is
\beq  \label{tut}   \F(\vj;\,X,Y',Y,Z',Z)\:=\:0.  \end{equation}
The dependence on $\,Q\,$ is trivial since it appears only in the combination
$\,\vj+\I^Q_{\ldots}\,.\;\;$ So $\,Q\,$ is not an independent variable.

\subsection{Re-parameterization of $\;\R$}

Let us introduce the new parameterization of the matrix (\ref{R-matrix}).
As we illustrated in \Fref{fumap}, in the auxiliary plane the mapping
$\Rop_{123}$ can be considered as a relative shift of three directed lines
$\Xl,\;\Yl,\;\Zl$ with respect to each other.
Now, for the given algebraic curve $\Gamma_g$ and $\vj\in\mathbf{C}^g$,
we introduce three pairs of points on $\Gamma_g$:
\beq X',\;X,\;Y',\;Y,\;Z',\;Z\:\in\:\Gamma_g. \label{XYZ}\end{equation}
Another point $Q\in\Gamma_g$ will just serve as a trivial normalization.
Then let
\beq        \label{repar}
\R\;=\;\R(p_1,p_2,p_3,p_4)\;\;\;\;\Longleftrightarrow\;\;\;\;\R\;=\;
\R(\vj;X',X;Y',Y;Z',Z)
\end{equation}
with, using the shorthand notations (\ref{defth}) and $p_j=(x_j,y_j)$:
\bea \fl  x_1&=&\!\! \frac{1}{q}\frac{\lc X,Z\rc}{\lc Y',Z\rc}
\frac{[X,Y]\:[Y,Z]}{[X,Y+Z]\:[Y]};\hs\,
y_1\;=\;\;\frac{e(Z,Z')\,e(X,Y')}{e(X,Z)\,e(Y',Z')}\;
\frac{[Z,Y]\:\theta(\vj+\I_Y^Q+\I_X^{Z'})}{[X,Y+Z]\:[Y]}\,;
\ny\\[1mm] 
\fl  x_2&=&\!\!\frac{\lc X',Z\rc}{\lc Y,Z\rc}
\frac{[X]\:[Y,X+Z]}{[X,Z]\:[Y,X]};\hs\hx
y_2\;=\; q\,\frac{e(Z,Z')\,e(X',Y)}{e(X',Z)\,e(Y,Z')}\;
\frac{[Z,X]\:\theta(\vj+\I_Y^Q+\I_X^{Z'})}{[X,Z]\:[Y,X]}\,;\ny\\[1mm] 
\fl  x_3&=&\!\! \frac{1}{q}\frac{\lc X,Z\rc}{\lc Y,Z\rc}
\frac{[X]\:[Y,Z]}{[X,Z]\:[Y]};\hs\hs\!
y_3\;=\; q\,\frac{e(Z,Z')\,e(X,Y)}{e(X,Z)\,e(Y,Z')}\;\;
\frac{[Z]\:\theta(\vj+\I_Y^Q+\I_X^{Z'})}{[X,Z]\:[Y]}\,;\ny\\[1mm] 
\fl  x_4&=&\!\! \frac{1}{q}\frac{\lc X',Z\rc}{\lc Y',Z\rc}
\frac{[X,Y]\:[Y,X+Z]}{[X,Y+Z]\:[Y,X]};\;\;\;\;
y_4=\frac{e(Z,Z')\,e(X',Y')}{e(X',Z)\,e(Y',Z')}\;
\frac{[Z,X\!+\!Y]\:\theta(\vj+\I_Y^Q+\I_X^{Z'})}{[X,Y\!+\!Z]\:[Y,X]}\,.
\ny\\[-2mm]  \fl &&\label{newp}
\end{eqnarray}
Actually, see (\ref{W-def}), for defining $\:\R_{123}\:$
we don't need the $\:y_i\:$ themselves but only the ratios \vspace*{-1mm}
\bea \fl \lefteqn{\frac{y_3}{y_1}=q\,\frac{\ee{Y}{Z'}}{\lc X,Y\rc}
\frac{[Z]\,[X,Y+Z]}{[X,Z]\,[Z,Y]};\hx
\frac{y_4}{y_1}=\!\frac{\ee{X}{Y'}}{\ee{X}{Z}}
\frac{[Z,X+Y]\,[Y]}{[Z,Y]\,[Y,X]}; \hx \frac{y_3}{y_2}=
\frac{\ee{X}{Z}}{\ee{X}{Y}}\;\frac{[Z]\,[Y,X]}{[Y]\,[Z,X]}}\ny\\ \fl && \eea
from which $\;e(Z,Z')\;$ and $\;\theta(\vj+\I_Y^Q+\I_X^{Z'})\;$ drop out.

Note that for this parameterization we used a generic algebraic curve and
generic points on this curve, and a generic point on its Jacobian, all in 
order to parameterize just three independent complex numbers $x_1,x_2,x_3$.
In (\ref{newp}) all $x_k,\;y_k$ are the periodical functions of
$\vj\in\mathrm{Jac}(\Gamma_g)\,$.

The parameterization (\ref{newp}) is suggested by a few assumptions:
First:
The prime forms shall appear in the $x_i$ only in the form of $N$-$th$
roots of (\ref{crossr}):
\beq \frac{\lc X,Z\rc}{\lc Y,Z\rc}\;=\;{\eeee{X}{Y}{Z}{Z'}}^{1/N}\!\!.
         \label{cross}\end{equation}
Second, considering (\ref{fuei}), we demand that the line-section
parameters $b_1,\;b_2,\;b_3,\;b_2'$ (sections of the line $\;\Xl\;$ in
\Fref{fumap}) should be proportional to $N$-$th$ roots of
theta functions of the form $\;[X,\ldots]\;$ defined in (\ref{defth}).
Analogously, the sections $c_1,\;\ldots,\;c_2'$ of the line $\Yl$ are assumed to
be proportional to $\:[Y,\ldots]\,.\;$ Finally, we consider that we want to
use Fay identities to provide the Fermat relations and the Hirota equations.

The merit of this parameterization will be seen in several places: when we
consider the transformed mappings $\R^{(2)},\:\ldots,\,\R^{(6)},\:$
when we re-write eqs.(\ref{hiro}) and when we construct composite weights in
\Sref{section3}.

We now must verify that (\ref{newp}), and its generalization to the other
 Fermat points in the MTE, give a consistent parameterization of the
relevant equations (\ref{ferma}), (\ref{xa}) and (\ref{hiro}).
We first check that (\ref{newp}) satisfies the Fermat relations
\beq   x_j^N\,+\,y_j^N\,=\,1.  \label{fer}\end{equation}
Indeed, these are true due to the Fay identity, which for
$\;A,B,C,D\in\Gamma_g\;$ we write as
\bea \fl -\;\lb A,C\rb\lb D,B\rb\;\Theta(\vj)\;\Theta(\V_B^A\,+\,\I_D^C)
&+&\lb A,B\rb\lb D,C\rb\;\Theta(\V_D^A)\;\Theta(\V_B^C)\ny\\ \fl
&&\hspace*{-2cm}+\;\lb A,D\rb\lb C,B\rb\;\Theta(\V_B^A)\;\Theta(\V_D^C)\;=\;0.
\label{fayy}\;\;\;\end{eqnarray}
For $j=1$ put in (\ref{fayy})
$\;(A,\:B,\:C,\:D\,)\rightarrow(Y',\:X,\:Z',\:Z\,)$ and
$\:\vj\rightarrow\vj'=\vj+\I_Y^Q\:,\;$ giving
\bea \fl\lb Z,X\rb\lb Y',Z'\rb(Y)(X,Y+Z)&-&\lb Z,Z'\rb\lb Y',X\rb(Z,Y)\:
\Theta(\vj'+\I_X^{Z'})\ny\\&& 
\fl \hs\hs\hs\hs-\;\lb Z',X\rb\lb Y',Z\rb(X,Y)(Y,Z)\:=\:0\,,\ny\eea
for $j=2$ put in (\ref{fayy})
$(A,\;B,\;C,\;D\,)\rightarrow (X',\;Y,\;Z',\;Z\,)$ and 
$\:\vj\rightarrow\vj''=\vj+\I_X^Q\:,\;$ giving
\bea\fl\lb Z',Y\rb\lb X',Z\rb(Y,X)(X,Z)&-&\lb Z,Y\rb\lb X',Z'\rb(X)(Y,X+Z)\ny\\
\fl &&\hs+\:\lb Z,Z'\rb\lb X',Y\rb(Z,X)\:\Theta(\vj''+\I_Y^{Z'})\,=\,0\,,\ny\eea
for $j=3$ put in (\ref{fayy}) $\;(A,\;B,\;C,\;D\,)\rightarrow
           (Z,\;X,\;Z',\;Y\,)\;\:$ and $\:\vj\rightarrow\vj^+=\vj+\I_Z^Q\:$:
\bea \fl\lefteqn{\lb X,Z'\rb\lb Y,Z\rb(X)(Y,Z)-\lb X,Z\rb\lb Y,Z'\rb(Y)(X,Z)}
\ny\\ \fl && \hspace*{63mm}
+\;\lb Z,Z'\rb\lb X,Y\rb(Z)\:\Theta(\vj+\I_Y^Q+\I_X^{Z'})\;=\;0\,,\ny \eea 
\begin{eqnarray}
\fl \lefteqn{\mbox{or}\hx\lb X,Z'\rb\lb Y,Z\rb\:\Theta(\vj^++\I_X^Z)\:
 \Theta(\vj^++\I_Y^{Z'})
\:-\:\lb X,Z\rb\lb Y,Z'\rb\,\Theta(\vj^++\I_Y^Z)\,\Theta(\vj^++\I_X^{Z'})}\ny\\
\fl \hs\hs\hs\hs
&+&\lb Z,Z'\rb\lb X,Y\rb\,\Theta(\vj^+)\,\Theta(\vj^++\I_X^Z+\I_Y^{Z'})\;=\;0\,.
\ny\end{eqnarray}
For $j=4$ put in (\ref{fayy}) $(A,\;B,\;C,\;D\,)\rightarrow
      (Y',\;Z,\;X',\;Z'\,),\;\;\vj\rightarrow\vj^*=\vj+\I_X^Q+\I_Y^{Z'}\:$.
\subsection{Line-section parameters and Hirota equations in terms of theta
functions}
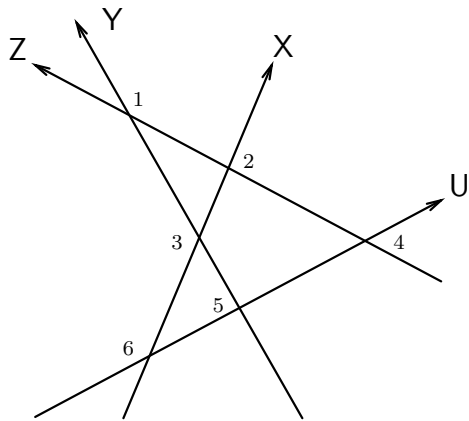
\begin{figure}[ht]
\setlength{\unitlength}{0.016mm}
\begin{center}
{\renewcommand{\dashlinestretch}{30}
\begin{picture}(3670,3450)(0,660)
\thicklines
\path(959,675)(2191,3613)
\path(2172,3491)(2191,3613)(2116,3513)
\path(3600,1813)(225,3613)
\path(345,3583)(225,3613)(316.8,3530)
\path(2445,675)(570,3965)
\path(654.502,3874.670)(570.000,3965.000)(602.052,3845.532)
\path(225,688)(3600,2488)
\path(3508.235,2405.059)(3600.000,2488.000)(3480.000,2458.000)
\put(3660,2520){\mk(0,0)[lb]{$\Ul$}}
\put(2200,3680){\mk(0,0)[lb]{$\Xl$}}
\put( 780,3900){\mk(0,0)[lb]{$\Yl$}}
\put(   0,3658){\mk(0,0)[lb]{$\Zl$}}
\put(1035,3270){\mk(0,0)[lb]{$\sk{1}$}}
\put(1950,2758){\mk(0,0)[lb]{$\sk{2}$}}
\put(1355,2083){\mk(0,0)[lb]{$\sk{3}$}}
\put(3200,2083){\mk(0,0)[lb]{$\sk{4}$}}
\put(1700,1563){\mk(0,0)[lb]{$\sk{5}$}}
\put( 950,1200){\mk(0,0)[lb]{$\sk{6}$}}
\end{picture}}
\caption{\footnotesize{Quadrangle
in the auxiliary plane formed by the directed
intersection lines of four oriented lattice planes. The six spaces
$\mathcal{V}$ in which the MTE operates are considered to be located at the
six intersection points.}} \label{quadro}\end{center}\end{figure}

\begin{center}
\begin{table}[ht]
\begin{center}\begin{tabular}{|ll|ll|}\hline &&&\\[-4mm]
$a_0$&$=\;[U]\lc U,X\rc\lc U,Y\rc\lc U,Z\rc$&
$b_1$&$=\;[X]\lc X,Y\!\rc\lc X,Z\!\rc\ee{U}{X}\!\!$\\[2mm]
$a_1$&$=\;[\,U,X]\lc U,Y\rc\lc U,Z\rc$ &
$b_2$&$=\;[X,Y]\lc X,Z\rc\ee{U}{X}$\\[3mm]
$a_1^\dg$&$=\;[\,U,Y]\lc U,Z\rc\lc U,X\rc$&
$b_2'$&$=\;[X,Z]\lc X,Y\rc\ee{U}{X}$\\[3mm]
$a_1'''=a_1^\dgh\!\!\!$&$=\;[\,U,Z]\lc U,X\rc\lc U,Y\rc$ &
$b_0$&$=\;[X,U]\lc X,Y\rc\lc X,Z\rc$\\[3mm]
$a_2$&$=\;[\,U,X+Y]\lc U,Z\rc$  &
$b_3   $&$=\;\:[X,Y+Z]\ee{U}{X}$\\[3mm]
$a_2^\dgg=a_2^T\!\!\!$&$=\;[\,U,Y+Z]\lc U,X\rc$  &
$b_1'''=b_1^t\!\!\!$&$=\;[X,U+Z]\lc X,Y\rc$\\[3mm]
$a_2''$&$=\;[\,U,X+Z]\lc U,Y\rc$ &
$b_1^\dg $&$=\;[X,U+Y]\lc X,Z\rc$\\[3mm]
$a_3$&$=\;[\,U,X+Y+Z]$&$b_2^\dgg=b_2^T\!\!\!$&$=\;\:[X,U+Y+Z]$\\[3mm]
\hline &&&\\[-4mm]
$c_2$&$    =\;[Y]\lc Y,Z\rc \ee{U}{Y}\:\ee{X}{Y}$&
$d_3$&$    =\;[Z]\ee{U}{Z}\:\ee{X}{Z}\:\ee{Y}{Z}$\\[2mm]
$c_1^\dg$&$=\;[Y,U]\lc Y,Z\rc\ee{X}{Y}$ &
$d_2'''=d_2^\dgh\!\!\!$&$=\;[Z,U]\ee{X}{Z}\:\ee{Y}{Z}$\\[3mm]
$c_3 $&$=\;[Y,Z]\ee{U}{Y}\:\ee{X}{Y}$&
$d_2'$&$=\:[Z,X]\ee{Y}{Z}\:\ee{U}{Z}$\\[3mm]
$c_1   $&$=\;[Y,X]\:\lc Y,Z\rc \ee{U}{Y}$ &
$d_2$&$=\;[Z,Y]\ee{U}{Z}\:\ee{X}{Z}$\\[3mm]
$c_0   $&$=\;[Y,U+X]\lc Y,Z\rc$  &
$d_1''=d_1^t\!\!\! $&$=\;[Z,U+X]\ee{Y}{Z}$\\[3mm]
$c_2^\dgh=c_2^T\!\!\!$&$=\;[Y,Z+U]\ee{X}{Y}$  &
$d_1   $&$=\;[Z,X+Y]\ee{U}{Z}$\\[3mm]
$c_2'  $&$=\;[Y,X+Z]\ee{U}{Y}$ &
$d_1^\dgg$&$=\;[Z,U+Y]\ee{X}{Z}$\\[3mm]
$c_1''=c_1^t\!\!\!$&$=\;[Y,Z+U+X]$&$d_0$&$=\;[Z,U+X+Y]$\\[1mm] \hline
\end{tabular}\end{center}
\caption{\footnotesize{The 32
line section parameters appearing in eqs.(\ref{hiro}),
expressed in terms of theta functions and prime factor ratios, using the
abbreviations (\ref{defth}). Observe that in the prime factor brackets 
$\:\lc\;,\;\rc\:$ the
points come always in the order $U,X,Y,Z$ (without and with primes).}}
\label{abcdt} \end{table}\end{center}
For writing the MTE in our new parameterization and to check (\ref{xa}) and
(\ref{hiro}), we consider three more
spaces $\mathcal{V}\:=\:\mathbb{C}^N\,,\:$ corresponding to the indices
$4,\,5,\,6\,.\:$ In \Fref{fumap} the first three spaces
were located at the intersection points of the lines $\Xl,\;\Yl,\;\Zl\,.$
To include the other three spaces, consider the "quadrangle" formed by 
{\it four} lines shown in \Fref{quadro}. Corresponding to the new line $\Ul$ 
we introduce another pair of points $U',U\in\Gamma_g.$
%
%

Looking at \Fref{quadro} we see that, instead of labeling the spaces by
the vertices $\,1,\ldots,\,6\,$ of the quadrangle, we can as well label them by
the pair of lines which intersect in these vertices, so identifying\\[-5mm]
\begin{equation}\fl
1\;\sim\;\Yl\Zl\:;\hq\; 2\;\sim\;\Xl\Zl\:;\hq\; 3\;\sim\;\Xl\Yl\:;\hq\;
4\;\sim\;\Ul\Zl\:;\hq\; 5\;\sim\;\Ul\Yl\:;\hq\; 6\;\sim\;\Ul\Xl\:.
\label{V-lines}\eeq
Note that the ordering of the lines is important for the identification 
(\ref{V-lines}): we shall chose the anti-clockwise orientation in
\Fref{quadro}, not the mirror reflected clockwise one.

Next we assume that we can write the ``coupling constants`` $\ka_j$ all in the 
form (\ref{cross}). Then from (\ref{V-lines}) it is suggestive to build e.g. 
$\ka_1$ from the points $Y',\;Y,\;Z',\;Z$ only, etc. and put (factors $q^{1/2}$ 
are inserted to produce correct signs when forming $N$-$th$ powers 
for (\ref{hiro})):
\bea \fl\ka_1  &=&q^{1/2}\frac{\lc Y',Z\rc}{\lc Y,Z\rc}\,;\hs
     \ka_2\;=\;q^{1/2}\frac{\lc X',Z\rc}{\lc X,Z\rc}\,;\hs
     \ka_3\;=\;q^{1/2}\frac{\lc X',Y\rc}{\lc X,Y\rc}\,;\ny\\[1mm]
     \fl \ka_4  &=&q^{1/2}\frac{\lc U',Z\rc}{\lc U,Z\rc}\,;\hs
     \ka_5\;=\;q^{1/2}\frac{\lc U',Y\rc}{\lc U,Y\rc}\,;\hs
     \ka_6\;=\;q^{1/2}\frac{\lc U',X\rc}{\lc U,X\rc}\,.
         \label{sumk}\eea
Now we consider (\ref{newp}) and (\ref{sumk}) and assume that the line-section
parameters $\:a_0,\:a_1,\ldots\:$ and $\:d_0,\:d_1,\ldots\:$ follow the same
scheme as postulated for $\:b_0,\ldots,\;c_0,\ldots$ above after (\ref{cross}):
$\:a_i\,\sim\:[U,\ldots]\,,\;\;d_i\,\sim\:[Z,\ldots]$. So equations (\ref{xa}) 
lead us to express all line-section parameters in terms of theta-functions as 
shown in \Tref{abcdt}.
We make ample use of the short-hand notations (\ref{defth}).

Apart from $Q$ which always comes with $\vj$, we use eight
 arbitrary points $X',X,Y',Y,$ $Z',Z,U',U\in \Gamma_g$.
 The $\ka_j$ may as well be written in terms of the bared brackets using
$\ee{A}{B'}\lc A,B\rc\:=\:\ee{A}{B}\lc A',B\rc\:.\;\;$
From \Tref{abcdt} we see that the $a_i$ don't depend on $U'$, the $b_i$
not on $X'$ etc.

Now, using the results of \Tref{abcdt} and (\ref{sumk}), we shall re-write all
the Hirota equations (\ref{hiro}) in terms of theta functions on $\Gamma_g$
and prime form cross ratios.
Not very surprisingly in view of \cite{krich,KWZ,ser-exact}, it turns out
that these all have the form of the double-Fay identity. Also, as expected
from \Fref{quadro}, and the meaning of the mappings as moving lines
within the quadrangle, the 24 equations follow from each other by a sequence of
simple substitutions.
Just inserting from \Tref{abcdt} and (\ref{sumk}), the first three equations 
of (\ref{hiro}) become, using the notation (\ref{DFay}) (recall that these are 
the equations (\ref{fli}) defining the functional mapping $\Ropf_{123}$):
\bea
\F(\vj;\,X,Y',Y,Z',Z)\:
          \frac{\lk\ee{U}{X}\ee{U}{Y}\ee{U}{Z}\ee{X}{Y}\ee{X}{Z}\rk^N}
                    {E(X,Y)\:E(X,Z)\:E(Y,Z)}&=&0\,;\ny\\
\F(\vj+\I^{Y'}_Y;\,Y',X',X,Z',Z)\:\;
               \frac{\lk\ee{U}{X}\ee{U}{Y}\ee{U}{Z}\rk^N}
                    {E(X,Z)\:E(X,Y')\:E(Y',Z)}\:&=&0\,;\ny\\
\F(\vj;\,Z,Y',Y,X',X)\:
           \frac{\lk\ee{U}{X}\ee{U}{Y}\ee{U}{Z}\lc X,Z\rc\lc Y,Z\rc\rk^N}
                    {E(X,Y)\:E(X,Z)\:E(Y,Z)}&=&0\,.\label{erdr}   \end{eqnarray}
The dependence on $U',\;U\,$ appears only in the factors on the right, not in
the $\F$. Assuming generic points $U',\:U,\ldots\:$ we conclude that the $\F$
must vanish and we combine the essential terms of (\ref{erdr}) into
\beq 
\FF(\vj;\,X',X,Y',Y,Z',Z)\;\stackrel{\textrm{\it\footnotesize def}}{=}\;\lk\!\!
\begin{array}{l}\F(\vj;\,X,Y',Y,Z',Z)\\[2mm]\F(\vj+\I^{Y'}_Y;\,Y',X',X,Z',Z)
\\[2mm]\F(\vj;\,Z,Y',Y,X',X)\end{array}\!\!\rk\!\!.   \end{equation}
Then the 24 Hirota equations (\ref{hiro}) which describe the functional mappings
take the form 
\beq   \begin{array}{ll}
\fl\ds \FF(\vj;\:X',X,Y',Y,Z',Z)\;=\;0\,;&\hs
\ds \FF(\vj+\I^{U'}_U;\:X',X,Y',Y,Z',Z)\;=\;0\,;\\[3mm]
\fl\ds \FF(\vj+\I^{X'}_X;\:U',U,Y',Y,Z',Z)\;=\;0\,;&\hs
\ds \FF(\vj;\:U',U,Y',Y,Z',Z)\;=\;0\,;\\[3mm]
\fl\ds \FF(\vj;\:U',U,X',X,Z',Z)\;=\;0\,;&\hs
\ds \FF(\vj+\I^{Y'}_Y;\:U',U,X',X,Z',Z)\;=\;0\,;\\[3mm]
\fl\ds \FF(\vj+\I^{Z'}_Z;\:U',U,X',X,Y',Y)\;=\;0\,;&\hs
\ds \FF(\vj;\:U',U,X',X,Y',Y)\;=\;0\,.\hs  \label{32eqs}
\end{array} \end{equation}
The equations in the left column of \eref{32eqs} precisely correspond to the
first twelve equations of \eref{hiro}. The last three equations of \eref{hiro}
are combined into the top equation of the right column of \eref{32eqs}.  
As already mentioned with (\ref{tea}), equations \eref{32eqs} together contain
the functional TE.

\subsection{Theta-parameterization of the simple modified tetrahedron equation}

Fianally, we use the parameterization (\ref{newp}) to re-write
the MTE. From (\ref{rff}) with (\ref{xa})
we find that the functional mapping just produces a permutation of the four
pairs of points $\:X,X',\ldots,U,U'\:\in\Gamma_g$,
together with shifts in the vector $\vj$. Of course, the result corresponds to
(\ref{32eqs}). Explicitly it is:
\begin{theorem}
The simple modified tetrahedron equation
may be parameterized in the terms of $\Gamma_g$,
$\vj\in\mathrm{Jac}(\Gamma_g)$ and
four pairs $X',X$, $Y',Y$, $Z',Z$, $U',U$ $\in$ $\Gamma_g$ by the definition
(\ref{repar}),(\ref{newp}),(\ref{xa}) as follows:
\beq   \begin{array}{ll}
\fl\ds \R^{(1)}=\:\R(\vj;\:X',X;Y',Y;Z',Z)\,;&
\ds \R^{(5)}=\:\R(\vj+\I^{U'}_U;\:X',X;Y',Y;Z',Z)\,;\\[3mm]
\fl\ds \R^{(2)}=\;\R(\vj+\I^{X'}_X;\:U',U;Y',Y;Z',Z)\,;&
\ds \R^{(6)}=\:\R(\vj;\:U',U;Y',Y;Z',Z)\,;\\[3mm]
\fl\ds \R^{(3)}=\:\R(\vj;\:U',U;X',X;Z',Z)\,;&
\ds \R^{(7)}=\:\R(\vj+\I^{Y'}_Y;\:U',U;X',X;Z',Z)\,;\\[3mm]
\fl\ds \R^{(4)}=\;\R(\vj+\I^{Z'}_Z;\:U',U;X',X;Y',Y)\,;&
\ds \R^{(8)}=\:\R(\vj;\:U',U;X',X;Y',Y)\,.
\end{array} \label{8par} \end{equation}
\end{theorem}
\textbf{Proof:} $\;$ Each $\,\R^{(j)}\,$ is determined by its three Fermat
points $\,x_1^{(j)},\:x_2^{(j)},\:x_3^{(j)}\,$. From \cite{gps} these points
are known in terms of the line-section parameters, see (\ref{xa}).
Inserting the theta-function expressions for the
line-sections from \Tref{abcdt} into \eref{xa} one finds that
the $\:x_i^{(j)}\:$ for $j=2,\ldots,8$ are obtained from those for $j=1\,,$
equations (\ref{newp}),
by the substitutions seen in (\ref{8par}). \hfill $\Box$\\[-3mm]

Using the correspondence between the labels $1,\ldots,6$ and the line labels
$\:\Ul,\:\Xl,\:\Yl,\:\Zl\:,$ given in \eref{V-lines},
$\;\:\R^{(1)}\,=\;\R_{123}$ may also be labeled as $\R^{\Xl\Yl\Zl}$ etc., and we
write the MTE as
\bea \lefteqn{
\R^{\Xl\Yl\Zl}(\vj)\;\R^{\Ul\Yl\Zl}(\vj+\I_X^{X'})\;\R^{\Ul\Xl\Zl}
(\vj)\;\R^{\Ul\Xl\Yl}(\vj+\I_Z^{Z'})}
\ny\\[2mm]&&\hspace{12mm}=\;\rho\;
\R^{\Ul\Xl\Yl}(\vj)\;\R^{\Ul\Xl\Zl}
(\vj+\I_Y^{Y'})\;\R^{\Ul\Yl\Zl}(\vj)\;\R^{\Xl\Yl\Zl}(\vj+\I_U^{U'}).
\label{RXYZ}\eea
This notation also indicates directly the three pairs of points on the
algebraic curve which parameterize the matrices $\R^{(j)}$ in (\ref{8par}).

In \cite{gps} we had shown that using simple re-scalings, out of the 24
line-section parameters listed in \Tref{abcdt} and the 6 parameters $\ka_1,
\ldots,\ka_6\,,$ only eight parameters are independent. Here we have
eight points on $\Gamma_g$ which can be chosen freely. In addition,
16 phases from taking the $N$-$th$ roots can be chosen freely. In terms of the
line-section parameters, the choice of the independent phases is the same as
the one explained in \cite{gps}.

\section{The fused vertex weight $\Rf$}
\label{section3}

\subsection{Open $\;N_1\times N_2 \times N_3\;$ box}

The natural graphical interpretation of the $\R$-matrix is a three dimensional
vertex, i.e. the intersection of three planes in 3-D space.
The six indices $\sig_j,\;\sig_j'$ are associated to the edges of the vertex,
recall \Fref{tri}.

The next step is the consideration of the intersection of three {\em sets} of
$N_1$, $N_2$ and $N_3$ parallel planes. This produces a finite open
cubic lattice of the size $N_1\times N_2\times N_3$. We call the corresponding
vertex object $\Rf$. It is the result of the fusion of elementary $\R$-matrices.

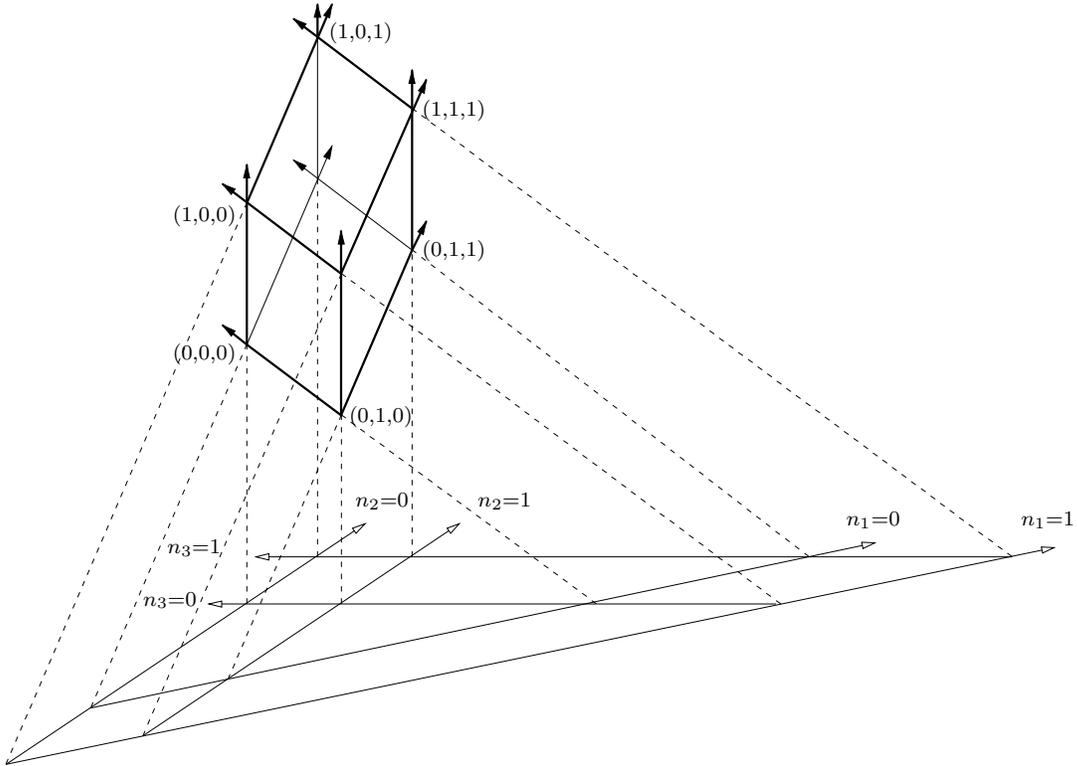
\begin{figure}[ht]
\begin{center}
\setlength{\unitlength}{0.00055in}
{\renewcommand{\dashlinestretch}{30}
\begin{picture}(10071,7294)(0,-10)
\thicklines
\path(3207,3342)(3207,4692)
\path(3207,3342)(3882,4917)
\path(3207,3342)(2307,4017)
\thinlines
\path(2982,6942)(2982,5592)
\dashline{60.000}(2982,5592)(2982,1992)
\dashline{60.000}(3882,4872)(3882,1992)
\dashline{60.000}(2307,4017)(822,552)
\dashline{60.000}(3207,3342)(2127,822)
\dashline{60.000}(3207,4692)(1317,282)
\dashline{60.000}(2307,5367)(12,12)
\dashline{60.000}(3207,3342)(3207,1542)
\dashline{60.000}(2307,4017)(2307,1542)
\dashline{60.000}(3207,3342)(5637,1542)
\path(12,12)(10002,2082)
\whiten\path(9890.583,2028.276)(10002.000,2082.000)
(9878.409,2087.028)(9890.583,2028.276)
\path(822,552)(8292,2127)
\whiten\path(8180.771,2072.888)(8292.000,2127.000)
(8168.392,2131.598)(8180.771,2072.888)
\dashline{60.000}(3207,4692)(7392,1542)
\dashline{60.000}(3882,4917)(7662,1992)
\thicklines
\path(3207,4692)(2082,5547)
\blacken\thinlines
\path(2195.692,5498.275)(2082.000,5547.000)
(2159.387,5450.505)(2195.692,5498.275)
\thicklines
\path(2307,4017)(2082,4197)
\blacken\thinlines
\path(2194.445,4145.463)(2082.000,4197.000)
(2156.963,4098.611)(2194.445,4145.463)
\thicklines
\path(3882,4917)(3882,6627)
\blacken\thinlines
\path(3912.000,6507.000)(3882.000,6627.000)
(3852.000,6507.000)(3912.000,6507.000)
\thicklines
\path(2307,4017)(2307,5727)
\blacken\thinlines
\path(2337.000,5607.000)(2307.000,5727.000)
(2277.000,5607.000)(2337.000,5607.000)
\thicklines
\path(3207,4692)(3207,5097)
\blacken\thinlines
\path(3237.000,4977.000)(3207.000,5097.000)
(3177.000,4977.000)(3237.000,4977.000)
\thicklines
\path(2982,6942)(2982,7257)
\blacken\thinlines
\path(3012.000,7137.000)(2982.000,7257.000)
(2952.000,7137.000)(3012.000,7137.000)
\thicklines
\path(2307,5367)(3117,7257)
\blacken\thinlines
\path(3097.304,7134.885)(3117.000,7257.000)
(3042.155,7158.520)(3097.304,7134.885)
\thicklines
\path(3207,4692)(4017,6537)
\blacken\thinlines
\path(3996.231,6415.063)(4017.000,6537.000)
(3941.292,6439.182)(3996.231,6415.063)
\thicklines
\path(3882,6267)(2757,7122)
\blacken\thinlines
\path(2870.692,7073.275)(2757.000,7122.000)
(2834.387,7025.505)(2870.692,7073.275)
\path(2307,4017)(3117,5907)
\blacken\path(3097.304,5784.885)(3117.000,5907.000)
(3042.155,5808.520)(3097.304,5784.885)
\path(3882,4917)(2757,5772)
\blacken\path(2870.692,5723.275)(2757.000,5772.000)
(2834.387,5675.505)(2870.692,5723.275)
\thicklines
\path(3882,4917)(4017,5187)
\blacken\thinlines
\path(3990.167,5066.252)(4017.000,5187.000)
(3936.502,5093.085)(3990.167,5066.252)
\path(12,12)(3432,2307)
\whiten\path(3349.073,2215.223)(3432.000,2307.000)
(3315.640,2265.045)(3349.073,2215.223)
\path(1317,282)(4332,2307)
\whiten\path(4249.110,2215.189)(4332.000,2307.000)
(4215.657,2264.997)(4249.110,2215.189)
\path(9597,1992)(2397,1992)
\whiten\path(2517.000,2022.000)(2397.000,1992.000)
(2517.000,1962.000)(2517.000,2022.000)
\path(7347,1542)(1947,1542)
\whiten\path(2067.000,1572.000)(1947.000,1542.000)
(2067.000,1512.000)(2067.000,1572.000)
\dashline{60.000}(3882,6267)(9597,1992)
\put(1320,1497){\mk(0,0)[lb]{$\sy{n_3=0}$}}
\put(1550,1992){\mk(0,0)[lb]{$\sy{n_3=1}$}}
\put(3342,2442){\mk(0,0)[lb]{$\sy{n_2=0}$}}
\put(4510,2442){\mk(0,0)[lb]{$\sy{n_2=1}$}}
\put(8022,2262){\mk(0,0)[lb]{$\sy{n_1=0}$}}
\put(9687,2262){\mk(0,0)[lb]{$\sy{n_1=1}$}}
\put(3090,6885){\mk(0,0)[lb]{$\sy{(1,0,1)}$}}
\put(3980,6145){\mk(0,0)[lb]{$\sy{(1,1,1)}$}}
\put(1600,3820){\mk(0,0)[lb]{$\sy{(0,0,0)}$}}
\put(3280,3230){\mk(0,0)[lb]{$\sy{(0,1,0)}$}}
\put(1600,5140){\mk(0,0)[lb]{$\sy{(1,0,0)}$}}
\put(3980,4810){\mk(0,0)[lb]{$\sy{(0,1,1)}$}}
\end{picture}
}
\end{center}
\caption{\footnotesize{Top:
3-dimensional view of the oriented cube $\:N_1=N_2=N_3=2\,$
(heavy lines) which is formed by six planes (indicated by dashed lines).
Bottom: the
horizontal auxiliary plane with the three pairs of lines arising from the
intersection of the three pairs of planes with the auxiliary plane. This is a
generalization of Figures \ref{tri} and \ref{fumap}:
If we consider e.g. the point
$(0,1,0)$ to correspond to the point $A$ of \Fref{tri}, then the inner
triangle in the auxiliary plane corresponds to the left shaded triangle of
\Fref{tri} and to the left part of \Fref{fumap}.
So the initial external
spin (Weyl) variables (\ref{inisig}) can be considered to sit at the three times
four intersection points of the auxiliary plane. In order to get a similar
picture for the final external variables we have to place the auxiliary plane
above the cube.}}
\label{cube} \end{figure}

The lattice is defined as in (\ref{cub}), but now $\:n_j=0,\ldots N_j-1\,.$
Let $\,\Rf_{123}\,$ be the matrix associated with the open cube (more precisely,
the open parallelepiped):
\bea\label{pf} \fl
\lefteqn{\ds \Rf_{123}\,\equiv\:\langle\vec\sig_1,\vec\sig_2,\vec\sig_3|\,\Rf\,
|\vec\sig_1',\vec\sig_2',\vec\sig_3'\rangle\:=\,\sum_{\{\sig\}}\,
\prod_{\pnt}\;
\langle\sig_{1,\pnt},\sig_{2,\pnt+\two},\sig_{3,\pnt}|\R_\pnt|
\sig_{1,\pnt+\one},\sig_{2,\pnt},\sig_{3,\pnt+\thr}\rangle\,.}\ny\\[-4mm]\fl &&
\eea
Here the six external multi-spin variables (i.e. the indices of
the matrix $\,\Rf_{123}$) are
associated with the six surfaces of the cube:
\beq \label{inisig} \fl
\ds \vec\sig_1=\;\{\sig_{1:0,n_2,n_3}\},\;\;
\vec\sig_2=\{\sig_{2:n_1,N_2,n_3}\},\;\;
\vec\sig_3=\{\sig_{3:n_1,n_2,0}\},\;\;n_j=0,...,N_j-1
\end{equation}
and \\[-5mm]
\beq       \label{outsig} \fl
\ds \vec\sig_1'=\{\sig_{1:N_1,n_2,n_3}\},\;\;
\vec\sig_2'=\{\sig_{2:n_1,0,n_3}\},\;\;
\vec\sig_3'=\{\sig_{3:n_1,n_2,N_3}\},\;\;n_j=0,...,N_j-1\,,
\end{equation}
and the summation in (\ref{pf}) is taken with respect to all internal indices
$\sig_{j,\pnt}$.
Anticipating what will be needed in (\ref{deriv}) in order to prove
that the fused weights satisfy a MTE of the same form as we had in
(\ref{RXYZ}), we use a reversed numbering for $\sig_2$ and $\sig'_2$,
so that the
"initial" external indices are $\sig_1=0,\;\sig_2=N_2,\;\sig_3=0$.
 This reversed numbering in the second space
is dictated also by our choice of the line orientations in the
lattice and, as consequence of this,
in the auxiliary plane (see \Fref{cube}).

In our next step we want to parameterize all $\R_{\pnt}$ in (\ref{pf}) such that
the fused weight $\Rf$ again satisfies a Modified Tetrahedron Equation. We shall
show that using a theta-function parameterization this is possible and the
$\,\Rf_{ijk}\,$ obtained will depend on $\:6(N_1+N_2+N_3)\:$ free parameters.

\begin{figure}
\begin{center}
\setlength{\unitlength}{0.15mm} 
\thicklines
\begin{picture}(450,450)
\put(25,25)
{\begin{picture}(400,400)
\put(50,0){\vector(1,2){200}}
\put(70,0){\vector(1,2){200}}
\put(30,0){\vector(1,2){200}}
\put(-37,-24){$n_2=0$}
\put( 65,-24){$n_2=2$}
\put(240,415){$Y',Y$}
\put(350,0){\vector(-1,2){200}}
\put(370,0){\vector(-1,2){200}}
\put(330,0){\vector(-1,2){200}}
\put(265,-24){$n_3=0$}
\put(365,-24){$n_3=2$}
\put(110,415){$Z',Z$}
\put(0,100){\vector(1,0){400}}
\put(0,120){\vector(1,0){400}}
\put(0,80){\vector(1,0){400}}
\put(-85,115){$n_1=0$}
\put(-85, 75){$n_1=2$}
\put(414,95){$X',X$}
\put(200,300){\circle{100}}
\put( 95,295){{\Large$\Vc_1$}}\put(130,300){\vector(1,0){30}}
\put(100,100){\circle{100}}
\put( 17,174){{\Large$\Vc_3$}}\put(40,160){\vector(1,-1){30}}
\put(300,100){\circle{100}}
\put(360,170){{\Large$\Vc_2$}}\put(360,160){\vector(-1,-1){30}}
\end{picture}}\end{picture}\end{center}
\caption{\footnotesize{Ordering
of the indices of the matrix $\Rf_{123}$, shown for the case
$\:N_1=N_2=N_3=3\:$ by drawing the auxiliary triangle in the auxiliary plane,
as in the bottom part of \Fref{cube}.}}
\label{fig-triangle}\end{figure}
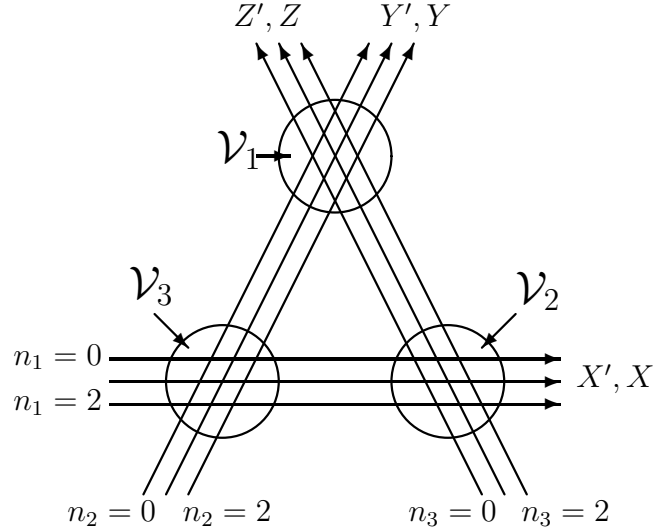

We use again the generic algebraic
curve $\Gamma_g$, and one vector $\vj\in\mathbb{C}^g$.
As in (\ref{repar}) each $\R_{\pnt}$ will depend on three pairs of points on
$\Gamma_g$, and to each $\R_{\pnt}$ we assign different three pairs:
\beq X_{n_1}',X_{n_1}^{},Y_{n_2}',Y_{n_2}^{},Z_{n_3}',Z_{n_3}^{},\hs
n_j=0,...,N_j-1.\end{equation}
However, the argument $\vj$ will be shifted for each $\R_{\pnt}$ by an amount
$\I_{\pnt}$ which depends on the points assigned to ''previous'' neighbors:
We define
\beq  \label{Rpt}\fl
\R_{\pnt}^{(123)}\;=\;\R(\vj+\I_{\pnt};X_{n_1}',X_{n_1}^{};Y_{n_2}',
Y_{n_2}^{};Z_{n_3}',Z_{n_3}^{})\;,\hs n_j=0,...,N_j-1,
\end{equation}
where
\beq  \label{Ipt}
\I_{\pnt}\;=\;\sum_{m_1=0}^{n_1-1}\,\I^{X_{m_1}'}_{X_{m_1}}
\,+\,\sum_{m_2=0}^{n_2-1}\,\I^{Y_{m_2}'}_{Y_{m_2}}
\,+\,\sum_{m_3=0}^{n_3-1}\,\I^{Z_{m_3}'}_{Z_{m_3}}\;.
\end{equation}
Now \eref{pf} and \eref{Rpt} define the matrix function
\beq\label{Zvj}  \Rf_{123}(\vj)\;=\;\Rf(\vj;X',X;Y',Y;Z',Z)
\end{equation}
where $X',X,Y',Y,Z',Z\;$ stand for the ordered lists of divisors,
\beq    \label{Xset}\fl
X=(X_0^{},X_1^{},...,X_{N_1-1}^{}),\;\;
X'=(X_0',X_1',...,X_{N_1-1}'),\;\;Y=(Y_0,Y_1,\ldots),\;\;\textrm{etc.}
\end{equation}
As to the index structure, recall (\ref{Vspace}),
\beq    \label{Zind}
\Rf_{123}\;\in\;\mathrm{End}\left(\Vc^{N_2N_3}\otimes
\Vc^{N_1N_3}\otimes\Vc^{N_1N_2}\right)\;,
\end{equation}
where in the same way as before we enumerate the number of
$\Vc^{N_jN_k}$ in the
tensor product (e.g (\ref{pf}) are the matrix elements of $\Rf_{123}$).
In \Fref{fig-triangle} we show the intersection lines of the planes of a
$N_1\times N_2\times N_3$ cube which appear in an auxiliary plane
(as in the bottom part of \Fref{cube}),
which intersects the ``initial'' edges corresponding
to (\ref{inisig}). On the section the $N_1+N_2+N_3$ planes become lines,
and the edges of the cubic lattice become the intersection points of
$N_1N_2+N_2N_3+N_1N_3$ lines in the auxiliary plane. The intersection points
are gathered into three sets $\Vc_1=\Vc^{N_2N_3}$ etc.,
and the index of $\Vc_j$ is the number of the corresponding $\Vc^{N_kN_l}$
in the tensor product in (\ref{Zind}).
\Fref{fig-triangle} is helpful to arrange the numbering in (\ref{Ipt}) and
the correct assignment of $X_{n_1}',X_{n_1}$, etc.

\subsection{The Modified Tetrahedron equation for the fused weights}

For writing the MTE, apart from the three pairs of sets $\;X',X;\:Y',Y;\:Z',Z\:$
of (\ref{Zvj}), (\ref{Xset}) we need a fourth pair
$$U\;=\;(U_0,\,U_1,\,...,U_{N_0-1}\,);\hs U'\;=
             \;(U_0',\,U_1',\,...,U_{N_0-1}'\;).$$
Applying the definition (\ref{pf}), in addition to (\ref{Zvj}) we construct
the matrices $$\Rf_{145}(\vj)=\Rf(\vj;U',U;Y',Y;Z',Z)\,;\hs
\Rf_{246}(\vj)=\Rf(\vj;U',U;X',X;Z',Z);$$
$$\Rf_{356}(\vj)=\Rf(\vj;U',U;X',X;Y',Y).$$
Their index structure is defined by
\beq\label{V-table} \begin{array}{l}
\Vc_1=\Vc^{N_2N_3}\;,\;\;
\Vc_2=\Vc^{N_3N_1}\;,\;\;
\Vc_3=\Vc^{N_1N_2}\;,\\
\Vc_4=\Vc^{N_0N_3}\;,\;\;
\Vc_5=\Vc^{N_0N_2}\;,\;\;
\Vc_6=\Vc^{N_0N_1}\;.
\end{array}
\end{equation} so that e.g. $\Rf_{145}$ is acting in a space of dimension
$N^{N_2N_3+N_0N_3+N_0N_2}\,.$
For $\,\Rf_{145}\,$ in analogy to the definitions (\ref{Rpt}),(\ref{Ipt})
one uses
\beq \R_{\pnt}^{(145)}\;=\;
  \R(\vj+\I_{\pnt};U_{n_0}',U_{n_0};Y_{n_2}',Y_{n_2};Z_{n_3}',Z_{n_3})
\end{equation}
with
\beq\I_{\pnt}\;=\;\sum_{m_0=0}^{n_0-1}\,\I^{U_{m_0}'}_{U_{m_0}}
            \,+\,\sum_{m_2=0}^{n_2-1}\,\I^{Y_{m_2}'}_{Y_{m_2}}
            \,+\,\sum_{m_3=0}^{n_3-1}\,\I^{Z_{m_3}'}_{Z_{m_3}}\,,
\end{equation}
similarly for $\Rf_{246}$ and $\Rf_{356}\,.$

\begin{theorem} The matrices $\Rf$ defined in (\ref{pf}) obey the modified
tetrahedron equation
\beq \label{zmte}
\begin{array}{l}
\displaystyle
\Rf_{123}(\vj)\;\Rf_{145}(\vj+\I_{X})\;\Rf_{246}(\vj)\;
    \Rf_{356}(\vj+\I_{Z})\;=\\ \\ \hspace{3cm}
\displaystyle\rho\;
\Rf_{356}(\vj)\;\Rf_{246}(\vj+\I_{Y})\;\Rf_{145}(\vj)\;
   \Rf_{123}(\vj+\I_{U})\end{array}
\end{equation}
where \\[-6mm]
\beq \begin{array}{l}
\I_{U}\;=\;\sum_{n_0=0}^{N_0-1}\,\I^{U_{n_0}'}_{U_{n_0}}\;;\hs\hs
\I_{X}\;=\;\sum_{n_1=0}^{N_1-1}\,\I^{X_{n_1}'}_{X_{n_1}}\;;\\
\\
\I_{Y}\;=\;\sum_{n_2=0}^{N_2-1}\,\I^{Y_{n_2}'}_{Y_{n_2}}\;;\hs\hs
\I_{Z}\;=\;\sum_{n_3=0}^{N_3-1}\,\I^{Z_{n_3}'}_{Z_{n_3}}\;.
\end{array} \label{Iuxyz}\end{equation}
\end{theorem}
\noindent
\textbf{Proof:} $\;\;$ The main content of this theorem is the appearance
of the specific set of shifts (\ref{Ipt}), (\ref{Iuxyz}).
For the proof it is convenient to
introduce some compact notations.
Instead of using the number labels for the $\Rf$ we shall use the labels
$\:\Ul,\,\Xl,\;\Yl,\;\Zl\:$ just as these were introduced in (\ref{V-lines}) and
(\ref{RXYZ}) for the single vertex matrices $\R$.
So, for the box $\Rf$-matrix and its sets of divisors we write,
\beq
\Rf_{123}(\vj)\;\Longrightarrow\;\Rf^{\Xl\Yl\Zl}(\vj);\hs
\Rf_{145}(\vj)\;\Longrightarrow\;\Rf^{\Ul\Yl\Zl}(\vj);\hs etc.
\end{equation}
In this short notation , formulas (\ref{pf}), (\ref{Rpt}) imply the definition
\beq \label{Ralg}
\Rf^{\Xl\Yl\Zl}(\vj)\;=\;
\prod_{n_1=0\uparrow N_1-1}\;\:\prod_{n_2=N_2-1\downarrow 0}\;\:
\prod_{n_3=0\uparrow N_3-1}\R^{X_{n_1}Y_{n_2}Z_{n_3}}(\vj+\I_{\pnt})
\end{equation}
where we use ordered products
\begin{equation}
\prod_{n_1=0\uparrow N_1-1}\mathfrak{f}_{n_1}\;=\;
\mathfrak{f}_0\mathfrak{f}_1\cdots\mathfrak{f}_{N_1-1}\;,\;\;
\prod_{n_2=N_2-1\downarrow 0}\mathfrak{f}_{n_2}\;=\;
\mathfrak{f}_{N_2-1}\cdots\mathfrak{f}_1\mathfrak{f}_{0}\;.
\end{equation}
For the triple $(X,Y,Z)\,,\;$ $\I_{\pnt}$ is given by (\ref{Ipt}).
Each $\R^{\Xl_{n_1}\Yl_{n_2}\Zl_{n_3}}$ 
acts non-trivially in only three of all the
spaces (\ref{V-table}) and the ordering is relevant just for neighboring indices
$X_{n_1}$ or $Y_{n_2}$ or $Z_{n_3}$.
The analogous modifications required to get the other matrices $\Rf_{145}$, etc.
should be evident.

Now we turn to the proof of the theorem which will be by
mathematical induction.
The theorem claims the validity of the MTE for arbitrary
$N_0,N_1,N_2,N_3$. For the initial point $N_0=N_1=N_2=N_3=1$ the MTE holds
because it is just (\ref{RXYZ}).
Then to prove the theorem, we reduce the MTE (\ref{zmte}) for some
$N_j$ to MTEs with $N_j'\le N_j$.
Thus one has four similar steps of the induction. Here we
consider the induction step for the $X$-direction, the other steps follow
analogously.

We split the list $X$ into two sublists of
length $N_1^{(1)}$ and $N_1^{(2)}=N_1-N_1^{(1)}$:
\beq
X^{(1)}=(X_0,X_1,...,X_{N_1^{(1)}-1})\;,\;\;\;
X^{(2)}=(X_{N_1^{(1)}},...,X_{N_1-1})\;,
\end{equation}
so that
$\I_{X^{(1)}}=\sum_{n_1=0}^{N_1^{(1)}-1}\I_{X_{n_1}}$ and
$\I_{X^{(2)}}=\I_{X}-\I_{X^{(1)}}$.
According to this splitting and due to the definition (\ref{Ralg})
\beq \label{X12dec}\begin{array}{l}
\Rf^{\Xl\Yl\Zl}(\vj)\;
=\;\Rf^{\Xl^{(1)}\Yl\Zl}
(\vj)\;\;\Rf^{\Xl^{(2)}\Yl\Zl}(\vj+\I_{X^{(1)}})\;,\\[2mm]
\Rf^{\Ul\Xl\Zl}(\vj)\;
=\;\Rf^{\Ul\Xl^{(2)}\Zl}(\vj+\I_{X^{(1)}})\;\Rf^{\Ul\Xl^{(1)}\Zl}(\vj)\;,\\[2mm]
\Rf^{\Ul\Xl\Yl}(\vj)\;
=\;\Rf^{\Ul\Xl^{(2)}\Yl}(\vj+\I_{X^{(1)}})\;\;\Rf^{\Ul\Xl^{(1)}\Yl}(\vj)\;.
\end{array}
\end{equation}
The meaning of notations $\Xl^{(1)}$ and $\Xl^{(2)}$ used in
(\ref{X12dec})  should be evident from (\ref{Ralg}).
Observe that because of the reverse numbering with respect to the middle
superscript of $\Rf$ in (\ref{Ralg}), the factors in the
latter two equations appear in reverse order.
Since $\:\Rf^{\Ul\Yl\Zl}(\vj)\:$ contains no $X$, neither as subscript nor 
in the argument, it will not be split. However, we have to put
\[\:\Rf^{\Ul\Yl\Zl}(\vj+\I_X)\:=\:\Rf^{\Ul\Yl\Zl}
                          (\vj+\I_{X^{(1)}}+\I_{X^{(2)}})\,.\]
Now substituting (\ref{X12dec}) to the LHS of (\ref{zmte}) written in our new
superscript notation, and abbreviating $\;\vj_1\;\equiv\;\vj+I_{X^{(1)}}\,,\;$
we get
\bea\fl\lefteqn{
\Rf^{\Xl\Yl\Zl}(\vj)\;\Rf^{\Ul\Yl\Zl}(\vj+\I_X)\;
\Rf^{\Ul\Xl\Zl}(\vj)\;\Rf^{\Ul\Xl\Yl}(\vj+\I_Z)}
\ny\\[3mm] \fl\hs
&=&\Rf^{\Xl^{(1)}\Yl\Zl}(\vj)
\;\Rf^{\Xl^{(2)}\Yl\Zl}(\vj+\I_{X^{(1)}})
\;\Rf^{\Ul\Yl\Zl}(\vj+\I_{X^{(1)}}+\I_{X^{(2)}})
\;\Rf^{\Ul\Xl^{(2)}\Zl}(\vj+\I_{X^{(1)}})\times\ny\\ \fl&& \hspace{4cm}\times
\;\Rf^{\Ul\Xl^{(1)}\Zl}(\vj)
\;\Rf^{\Ul\Xl^{(2)}\Yl}(\vj+\I_{X^{(1)}}+\I_Z)
\;\Rf^{\Ul\Xl^{(1)}\Yl}(\vj+\I_Z)\ny\\[3mm]\fl
&=&\Rf^{\Xl^{(1)}\Yl\Zl}(\vj)\;\left[
\Rf^{\Xl^{(2)}\Yl\Zl}(\vj_1)
\:\Rf^{\Ul\Yl\Zl}(\vj_1+\I_{X^{(2)}})
\:\Rf^{\Ul\Xl^{(2)}\Zl}(\vj_1)\:\Rf^{\Ul\Xl^{(2)}\Yl}(\vj_1+\I_Z)\:\right]\times
\ny\\  \fl&&\hspace{6cm}\times
\Rf^{\Ul\Xl^{(1)}\Zl}(\vj)
\Rf^{\Ul\Xl^{(1)}\Yl}(\vj+\I_Z)\ny\\[3mm]\fl
&=&\Rf^{\Xl^{(1)}\Yl\Zl}(\vj)\;\left[
\Rf^{\Ul\Xl^{(2)}\Yl}(\vj_1)
\;\:\Rf^{\Ul\Xl^{(2)}\Zl}(\vj_1+\I_Y)
\;\:\Rf^{\Ul\Yl\Zl}(\vj_1)\;\:\Rf^{\Xl^{(2)}\Yl\Zl}(\vj_1+\I_U)\:\right]\times
\ny\\ \fl &&\hspace{6cm}\times
\Rf^{\Ul\Xl^{(1)}\Zl}(\vj)
\Rf^{\Ul\Xl^{(1)}\Yl}(\vj+\I_Z)\ny\\ \fl
&=&\Rf^{\Ul\Xl^{(2)}\Yl}(\vj_1)
\:\Rf^{\Ul\Xl^{(2)}\Zl}(\vj_1+\I_Y)\times\ny\\ \fl&&\hspace{2mm}\times\left[
\Rf^{\Xl^{(1)}\Yl\Zl}(\vj)
\:\Rf^{\Ul\Yl\Zl}(\vj+X^{(1)})\:\Rf^{\Ul\Xl^{(1)}\Zl}(\vj)
\:\Rf^{\Ul\Xl^{(1)}\Yl}(\vj+\I_Z)\right]
\Rf^{\Xl^{(2)}\Yl\Zl}(\vj_1+\I_U)\ny\\[3mm] \fl
&=&\Rf^{\Ul\Xl^{(2)}\Yl}(\vj_1)
\;\Rf^{\Ul\Xl^{(2)}\Zl}(\vj_1+\I_Y)\times\ny\\ \fl
&&\hspace{2mm}\times\left[\Rf^{\Ul\Xl^{(1)}\Yl}(\vj)
\:\Rf^{\Ul\Xl^{(1)}\Zl}(\vj+\I_Y)\:\Rf^{\Ul\Yl\Zl}(\vj)
\:\Rf^{\Xl^{(1)}\Yl\Zl}(\vj+\I_U)\right]\:
\Rf^{\Xl^{(2)}\Yl\Zl}(\vj_1+\I_U)\ny\\[3mm] \fl
&=&\Rf^{\Ul\Xl^{(2)}\Yl}(\vj+\I_{X^{(1)}})
\:\Rf^{\Ul\Xl^{(1)}\Yl}(\vj)\
\:\Rf^{\Ul\Xl^{(2)}\Zl}(\vj+\I_{X^{(1)}}+\I_Y)
\:\Rf^{\Ul\Xl^{(1)}\Zl}(\vj+\I_Y)\times \ny\\ \fl
&&\hspace{4cm}\times\Rf^{\Ul\Yl\Zl}(\vj)
\:\Rf^{\Xl^{(1)}\Yl\Zl}(\vj+\I_U)
\:\Rf^{\Xl^{(2)}\Yl\Zl}(\vj+\I_{X^{(1)}}+\I_U)\ny\\[3mm]\fl
&=&\Rf^{\Ul\Xl\Yl}(\vj)\;\Rf^{\Ul\Xl\Zl}(\vj+\I_Y)\;
\Rf^{\Ul\Yl\Zl}(\vj)\;\Rf^{\Xl\Yl\Zl}(\vj+\I_U)\,.
\label{deriv}\eea
From the third to the fourth line of (\ref{deriv})
within the inserted brackets we used the MTE for
the smaller set $(U,X^{(2)},Y,Z)$. In order to be
able to isolate the terms containing $X^{(2)}$ the order of factors in the
second line of (\ref{X12dec}), which was used in the first step, is crucial.
Going from the fifth to the sixth line in the brackets we used the MTE for
$(U,X^{(1)},Y,Z)\,.$ In the other steps of (\ref{deriv}) we just commuted
or combined various terms. The last step is made possible by the "reverse"
order of factors in the last line of (\ref{X12dec}).\\[-1mm]
\hfill $\Box$

\section{Special cases: Solving the tetrahedron equation}

\subsection{Compact algebraic curve}

Let now $N_0=N_1=N_2=N_3\,=\,M\,$.
Starting from the generic parameterization of the MTE (\ref{zmte}),
the usual tetrahedron equation is obtained if
\beq  \Rf(\vj)\;\equiv\;\Rf(\vj+\I)\hs\;\;\textrm{and}\hs
  \;\rho\;=\;1\;.\end{equation}
This is the case when
\beq   \label{zero}
\I_{U}=\I_{X}=\I_{Y}=\I_{Z}\;=\;0 \hs\textrm{on}\hs
         \mathrm{Jac}(\Gamma_g)\;,   \end{equation}
and the ratios of $\theta$-functions (\ref{etheta}) are periodical.

According to Abel's theorem,
\eref{zero} means that there are four meromorphic functions
$u,\:x,\:y,\:z\:$ on  $\Gamma_g$ with the divisors
\bea       \label{xyzu}  \fl
\lefteqn{(u)=\!\sum_{n_0=0}^{M-1} U_{n_0}'-\!U_{n_0}^{},\;\;\,
(x)=\!\sum_{n_1=0}^{M-1} X_{n_1}'-\!X_{n_1},\;\;\,
(y)=\!\sum_{n_2=0}^{M-1} Y_{n_2}'-\!Y_{n_2},\;\;\,
(z)=\!\sum_{n_3=0}^{M-1} Z_{n_3}'-\!Z_{n_3}.}\ny\\[-2mm]\fl &&
\eea
As it is well known (see e.g. Theorem 10-23 of \cite{springer}),
conditions (\ref{zero}) are a strong restriction for the type of $\Gamma_g$:
$\Gamma_g$ is the algebraic curve given by the polynomial
equation
\beq \label{Polynom}
P(x,y)\;\stackrel{\textrm{\it\footnotesize def}}{=}\;\sum_{a,b=0}^{M}
               \,x^a\,y^b\,p_{a,b}\;=\;0\;. \end{equation}
The choice of any pair of $\,x,\,y,\,z,\,u\,$ produces an
equivalent polynomial equation.
The form of the polynomial $P(x,y)$ provides the restriction for the genus,
\beq   g\;\leq\;(M-1)^2\;.  \end{equation}
Thus we come to
\begin{theorem} \label{Thdrei}
Let $\Gamma_g$ be the compact algebraic curve defined by
polynomial equation (\ref{Polynom}).
Let four sets of $U_{n_0}',U_{n_0}^{}$,
 $X_{n_1}',X_{n_1}^{}$, $Y_{n_2}',Y_{n_2}^{}$ and
$Z_{n_3}',Z_{n_3}^{}$,  $n_k=0,...,M-1$,
are the divisors of four meromorphic
functions $u,x,y,z$ on $\Gamma_g$. Then the tetrahedron equation is satisfied
\beq\label{te}  \begin{array}{l}
\ds\hspace{-1cm}
\Rf_{123}(x,y,z)\:\Rf_{145}(u,y,z)\:\Rf_{246}(u,x,z)\:\Rf_{356}(u,x,y)\;=\\[2mm]
 \ds\hs\Rf_{356}(u,x,y)\:\Rf_{246}(u,x,z)\:\Rf_{145}(u,y,z)\:\Rf_{123}(x,y,z)\;,
\end{array}
\end{equation}
where four matrices are the same matrix function of different arguments,
\begin{equation}
\Rf(x,y,z)\;=\;\Rf(\vj;X',X;Y',Y;Z',Z)\;\;\;\textrm{etc.}
\end{equation}
defined via (\ref{pf}),(\ref{Rpt}),(\ref{Zvj}) and (\ref{xyzu}).
\end{theorem}
According to the conventional terminology,
one may say that $u,x,y,z$ are the spectral parameters, the moduli
of $\Gamma_g$ are the moduli of the tetrahedron equation, and vector
$\vj$ is a kind of deformation parameter.

Theorem \ref{Thdrei} may be also be proved differently, without mentioning
the simple MTE. In this alternative approach one considers the auxiliary
linear problem for the whole box and the corresponding mappings.
See e.g. \cite{s-qem,s-kiev} for the description of this method and
\cite{sp-cl} for the parameterization of the classical equations of motion.
Remarkably, in this alternative way the spectral curve (\ref{Polynom})
appears naturally from the linear problem.

\subsection{Simple tetrahedron equation for ZBB-case recovered}

In section 1 we discussed the ZBB-model and its $\,\R$-matrix \eref{R-matrix}.
We now show how our scheme contains this case.
Formally ZBB's tetrahedron equation corresponds to (\ref{te}) with $M=1$.
It gives $g=0$, i.e. the spectral curve is the sphere with
\begin{equation}
E(X,Y)\;=\;\frac{X-Y}{\sqrt{dX\,dY}}\:,\hs\hs
\Theta(\textrm{ }\;)\equiv 1\;.
\end{equation}
Here the formal theta-function has no argument since the Jacobian 
is $0$-dimensional.
Conditions (\ref{zero}) therefore are out of use, and the parameterization
(\ref{repar}),(\ref{newp}) contains the $N$-$th$ roots of the cross-ratios like
$\ds \frac{(X-Z)(X'-Z')}{(X'-Z)(X-Z')}$. One may show that
the number of independent cross-ratios is the number of variables $X,X',...$
\emph{minus} three. Therefore the single $\R$-matrix contains
$6-3=3$ independent complex parameters (as it should be),
and the simple tetrahedron equation
contains $8-3=5$ independent complex parameters (again as it should be).
It means that the tetrahedral condition, which appears in Zamolodchikov's
parameterization of $\R$ in the terms of spherical triangles, is taken
into account automatically. Moreover,
the parameterization with the help of the cross-ratios takes
automatically into account the geometric structure of any set of the
planes in three dimensional Euclidean space.
The parameterization of the inhomogeneous Zamolodchikov-Bazhanov-Baxter
model in the terms of cross-ratios corresponding to the $\,g=0\,$ limit of
(\ref{repar}),(\ref{newp}),(\ref{Rpt}) has already been used in \cite{svan}.

\subsection{Chessboard model}

Previously derived ``chessboard models'' of the lattice statistical mechanics
based on the modified tetrahedron equation \cite{mss-mte,bms-mte}
are of course related to our considerations. The term ``chessboard''
appeared as
the visual interpretation of the cubic lattice with $M=2$ to be homogeneous. It
means that the cubic lattice consists of eight different types of
vertices (i.e. eight
different types of the Boltzmann weights) -- a kind of three dimensional
analogue of the chess board with eight different colors of the cells.

The models described in \cite{mss-mte,bms-mte} are at the first
the so-called IRC-type models, but with the help of
vertex-IRC correspondence \cite{psi} one may construct their
vertex reformulation. Thus the model implicitly described
in \cite{bms-mte} is equivalent to $M=2$, $g=1$ of our scheme.
For $g=1$ the curve and its Jacobian are isomorphic, so that without loss of
generality one may chose
\begin{equation}
\I^X_Y\;=\Xt-\Yt\;\in\;\mathbb{C}/\mathbb{Z}+\mathbb{Z}\tau\;.
\end{equation}
Further, one may use $\theta_1$
\begin{equation}
\Theta(v)\;=\;
\theta_1(v,\tau)\;\equiv\;
\sum_{n=-\infty}^{\infty}\; e^{i\pi\tau(n+1/2)^2+2i \pi
  (v+1/2)(n+1/2)}
\end{equation}
as basic theta-function,
and  $E(X,Y)\sim\theta_1(\Xt-\Yt)$ as the prime form. This formulas simplify the
definitions (\ref{etheta}). Periodicity conditions (\ref{xyzu}) may be chosen
\bea \label{xyzu2}
\lefteqn{\Xt_0'-\Xt_0+\Xt_1'-\Xt_1\;=\;\Yt_0'-\Yt_0+\Yt_1'-\Yt_1}
  \ny\\ \hspace*{15mm}=\;\Zt_0'-\Zt_0+\Zt_1'-\Zt_1\;=\;
\Ut_0'-\Ut_0+\Ut_1'-\Ut_1\;=\;1\,.\eea
Note, the $1$ in the right hand side of \eref{xyzu2} is equivalent to 
$\tau$ (due to the Jacobi transform), while $0$ instead of $1$ in the right 
hand side of \eref{xyzu2} gives a trivial model.

The model explicitly described before in \cite{mss-mte} corresponds
to $M=2$, $g=1$, $\Xt_0^{}=\Xt_1^{}$, $\Xt_0'=\Xt_1'$, $\Yt_0^{}=\Yt_1^{}$,
$\Yt_0'=\Yt_1'$,
$\Zt_0^{}=\Zt_1^{}$, $\Zt_0'=\Zt_1'$ etc. with the condition \eref{xyzu2}.
This choice leads to the identification of the parameters in \eref{pf}
\begin{equation}
\R_{\pnt}=\R_{\pnt+\one+\two}=\R_{\pnt+\one+\thr}=\R_{\pnt+\two+\thr}\;,
\end{equation}
so the cells of this three dimensional chess board have only two ``colors''.

Note that the vertex-IRC duality is not exact because it changes the boundary
conditions.


\section{Conclusions}

We considered a large class of integrable 3-D lattice models which have
Weyl variables at $N$-$th$ root of unity as dynamic variables.
We have shown how the Boltzmann weights can be conveniently parameterized in
terms of $N$-$th$ roots of theta functions on a Jacobian of a compact
Riemann surface. The Fermat relations of the points determining the Boltzmann
weights are simple Fay identities and the classical equations determining the
scalar parameters are a consequence of a double Fay identity. In the modified
tetrahedron equation we have four pairs of arbitrary points on the Riemann
surface in simple permuted combinations.

This parameterization allows a compact formulation of the rules to form
fused Boltzmann weights $\Rf\in\mathrm{End}\: \mathbb{C}^{3N M^2}$ which are
the partition functions of open boxes of arbitrary size.
The $\Rf$ obey the modified tetrahedron equation and are again parameterized
terms on $N$-th roots of theta-functions on the
Jacobian of a genus $g=(M-1)^2$ compact Riemann surface $\Gamma_g$.
The spectral parameters of the vertex weight $\Rf$
are three meromorphic functions on the spectral curve $\Gamma_g$.
For the case that the Jacobi transforms become trivial the $\Rf$ obey the simple
tetrahedron equation. The Zamolodchikov-Baxter-Bazhanov model and the
Chessboard model are obtained as special cases.

So, the scheme discussed here
contains and generalizes many known 3-D integrable models. The hope is that
the framework is now sufficiently general to contain physically interesting
models with a non-trivial phase structure. However, to get information on
partition functions, either analytically or approximately,
is still a very difficult open problem. There is no way known to generalize
Baxter's special method \cite{bax-part} by which he obtained the partition
function of the ZBB-model.

\section*{Acknowledgments}
S.P. is grateful to Bonn and Angers Universities, where the work
was done partially. S.S. thanks the Max-Planck-Institut f\"ur
Mathematik and Bonn University. This work has been supported in
part by the contract INTAS OPEN 00-00055 and by the
Heisenberg-Landau program HLP-2002-11. S.P.'s work was supported
in part by the grants CRDF RM1-2334-MO-02, RFBR 03-02-17373 and
the grant for support of scientific schools 1999.2003.2. S.S.'s
work was supported in part by the grants CRDF RM1-2334-MO-02 and
RFBR 01-01-00201.

\section*{References}
\bibliographystyle{amsplain}

\end{document}